\documentclass[floats,floatfix,amssymb,prd,twocolumn,superscriptaddress,nofootinbib,longbibliography]{revtex4-1}

\AtBeginDocument{%
    \newwrite\bibnotes
    \def\bibnotesext{Notes.bib}
    \immediate\openout\bibnotes=\jobname\bibnotesext
    \immediate\write\bibnotes{@CONTROL{REVTEX41Control}}
    \immediate\write\bibnotes{@CONTROL{%
    apsrev41Control,author="08",editor="1",pages="1",title="0",year="1"}}
     \if@filesw
     \immediate\write\@auxout{\string\citation{apsrev41Control}}%
    \fi
}%

\usepackage{graphicx,amssymb,amsmath,amsthm,amsfonts,epsfig,epsf}

\usepackage[colorlinks=true,citecolor=purple,linktocpage]{hyperref}
\usepackage[usenames]{xcolor}
\usepackage{epstopdf}
\usepackage{bm}
\usepackage{dcolumn}
\usepackage{rotating}
\usepackage{hyperref}
\usepackage{xcolor}
\usepackage{longtable}

\usepackage{enumerate}
\usepackage{tensor,multirow}
\usepackage{url}
\usepackage{physunits} 
\usepackage{mathrsfs} 
\usepackage[normalem]{ulem}
\usepackage[export]{adjustbox}
\usepackage{braket}
\usepackage{dsfont}
\usepackage{enumitem}

\usepackage{booktabs}
\usepackage{multirow}

\allowdisplaybreaks 

\DeclareMathOperator{\diag}{diag} 

\newcommand{\dd}{\mathop{}\!\mathrm{d}}

\newcommand{\Mbh}{M_{\text{\tiny BH}}}

\newcommand{\Proca}{\text{Proca}}
\newcommand{\Mnps}{M_{\text{\tiny NPS}}}

\usepackage{pifont}

\begin{document}

\title{Extreme mass-ratio inspirals into Newtonian Proca stars}

\author{João Bernardo Silva}
\email{joaobernardosilva@tecnico.ulisboa.pt}
\affiliation{CENTRA, Departamento de F\'{\i}sica, Instituto Superior T\'ecnico -- IST, Universidade de Lisboa -- UL, Avenida Rovisco Pais 1, 1049-001 Lisboa, Portugal}
\author{Richard Brito}
\email{richard.brito@tecnico.ulisboa.pt}
\affiliation{CENTRA, Departamento de F\'{\i}sica, Instituto Superior T\'ecnico -- IST, Universidade de Lisboa -- UL, Avenida Rovisco Pais 1, 1049-001 Lisboa, Portugal}

\begin{abstract}
Massive bosonic fields can form self-gravitating solitonic structures, which for vector fields are known as Proca stars. For ultralight fields, these structures can describe the cores of dark matter haloes surrounding the supermassive black holes at the center of galaxies. It has been argued that future gravitational-wave detectors might be able to probe the properties of dark matter structures. However, most of the analyses considering ultralight dark matter have focused on massive scalar fields. In this work, we study how Proca stars respond to a perturbing small object inspiralling in their interior, in the Newtonian limit. We consider both Newtonian spherically symmetric Proca stars and the nonspherically symmetric ground-state solutions. We compute the total energy lost by the orbiting object and compare it to the case where the bosonic star is composed of a scalar field. Our results show that, at the Newtonian level, the energy lost by the object in a Proca star ground state is similar to that in its scalar field counterpart, with relative differences of at most $\sim 20\%$, while the maximum orbital energy loss rate in the ground-state configuration can be one to two orders of magnitude larger than in the spherically symmetric Proca star, when a central parasitic black hole is present. Our work motivates the need to study extreme mass-ratio inspirals into Proca stars using a fully relativistic setup, where differences between the Proca and scalar boson star ground states are expected to become more significant.
\end{abstract}

\date{\today}

\maketitle

\section{Introduction}

In the standard lambda cold dark matter ($\Lambda$CDM) model, dark matter (DM) accounts for $\sim 26\%$ of the Universe's energy density \cite{Planck2018}, yet its nature remains elusive. Of the many DM candidates proposed to date (see \cite{Bertone_WIMP,Bozorgnia:2024} for reviews), weakly interacting massive particles (WIMPs) remain among the most prominent \cite{WIMP_weinberg,WIMP_HUT,Leane:2018k_WIMP,Bottaro:2021_WIMP}, with masses typically $\gtrsim \mathrm{GeV}$, such as the lightest neutralino in the minimal supersymmetric extension of the Standard Model \cite{Jungman_1996,Roszkowski_neutralino,Bramante_neutralino}. However, while the $\Lambda$CDM model successfully describes structure formation on large scales, DM-only simulations exhibit long-standing tensions with observations on galactic scales ($\lesssim \mathrm{kpc}$), notably the cusp-core and missing-satellites problems \cite{Bullock:2017,Flores_1994,Klypin_1999}. Although some of these discrepancies may be alleviated, at least in part, by baryonic effects and observational incompleteness \cite{Bullock:2017,Adhikari:2022,Brooks_2013,jeon2025}, they have nevertheless motivated an interest in alternative DM models. 

One such class of alternatives is ultralight DM, composed of nonrelativistic bosonic fields with masses $ \ll 1~\mathrm{eV}$ \cite{Ferreira:2020}. Possible candidates include the QCD axion, originally introduced in the Peccei-Quinn solution to the strong-$CP$ problem \cite{Peccei_1977}, axionlike particles \cite{Choi:2020,Irastorza:2018} arising naturally in beyond-the-Standard-Model scenarios and string compactifications \cite{Arvanitaki_2010,witten_2006,Alexander:2024}, ``dark photons" associated with hidden $U(1)$ sectors \cite{Holdom1985,Fabbrichesi:2021DarkPhoton}, and even possibly massive tensor fields appearing in modified-gravity frameworks such as massive gravity and bimetric theory \cite{Schmidt-May:2015vnx,deRham:2014zqa,Marzola:2017lbt}. 

Ultralight bosonic fields can form self-gravitating macroscopic Bose-Einstein condensates, where gravitational attraction is balanced by quantum pressure, giving rise to solitonic configurations. In particular, we focus on the fuzzy dark matter (FDM) regime \cite{Eberhardt_ferreira:2025,Hu:2000ke,Capanelli:2025ykg}, with masses typically $m\sim10^{-22}-10^{-19}~\mathrm{eV}$. For these ultralight masses, the de Broglie wavelength can be of the order of kiloparsecs \cite{Hu:2000ke,Su_rez_2013,Li_2014}, leading to wavelike behavior on galactic scales \cite{Hui:2021}. This may alleviate some of the small-scale structure tensions associated with the $\Lambda$CDM model while preserving its successful large-scale phenomenology. The resulting solitonic configurations have been studied both as models for the cores of galactic DM haloes \cite{Lee_1996,Lee_2018,mourelle2025,Broadhurst_2020,mourelle_2025_after} and as exotic compact objects (ECOs) capable of mimicking black holes (BHs) \cite{Macedo_2013,Guzm_n_2009,Herdeiro_2021,Sanchis-Gual:2018,Marks:2025,Macedo:2026}. Ultralight bosonic fields may also play a role in other astrophysical contexts, such as bosonic clouds around spinning BHs formed via superradiance \cite{Brito:2015oca,Herdeiro:2014goa,Herdeiro:2016tmi}. These systems can leave imprints on gravitational-wave (GW) signals, making GW observations a powerful complementary probe of ultralight DM \cite{Bertone:2019irm,Miller:2025yyx}.

In the $\sim 10 - 10^3\,\mathrm{Hz}$ band, the ground-based LIGO-Virgo-KAGRA (LVK) detector network has established GW astronomy as a probe of compact binaries in the relativistic strong-field regime \cite{LIGOScientific:2025slb,LIGOScientific:2026qni}.  
Future space-based detectors \cite{Taiji:2018tsw,TianQin:2015yph,LISA:2024hlh}, such as the Laser Interferometer Space Antenna (LISA) \cite{LISA:2024hlh}, will extend GW observations to the millihertz band, where extreme mass-ratio inspirals (EMRIs) are among the key sources \cite{Berry:2019wgg,Babak_2017,Speri:2026ade}. These are binary systems in which a stellar-mass compact object of mass $m_p$, the secondary, inspirals into a much more massive central body of mass $M$, the primary, with mass-ratio $\varepsilon = m_p / M \lesssim 10^{-4}$. They are expected to form when compact stellar remnants, such as BHs and neutron stars, spiral into massive black holes (MBHs) in galactic nuclei, typically with $M \sim 10^4 - 10^7 \,M_\odot$ \cite{Babak_2017,Berry:2019wgg}. Because of their extreme mass-ratios, EMRIs evolve slowly, producing long-lived GW signals that can remain in band for months to years and accumulate $\sim 10^4 - 10^5$ cycles. These signals encode detailed information about the spacetime geometry of the primary, making EMRIs powerful probes of general relativity and possible deviations from it \cite{Barack_2007,Sopuerta:2009iy,Pani_2011,Gair:2011ym,Cardenas-Avendano:2024mqp,Kuntz:2020yow,Zhang:2022rfr,Speri:2024qak}, as well as of ECOs \cite{Macedo_2013,Datta:2019epe,Pani:2019cyc,Annulli:2020l,Maggio:2021,Destounis:2023khj,Zi:2024itp,Nair:2025anr,Macedo:2026} and astrophysical environments~\cite{Barausse:2014tra}, including those sourced by ultralight bosonic fields \cite{Hannuksela:2018izj,Baumann:2019ztm,Annulli:2020l, Tomaselli:2023ysb,Brito:2023,Duque_2024,Li:2025ffh,Karmakar:2025drp,Dyson:2025,Xu:2026aic,Xu:2026cky}.

Among the nonvacuum compact objects that may serve as EMRI primaries are self-gravitating solitonic configurations formed by massive complex bosonic fields, known as boson stars (BSs) for scalar fields \cite{kaup1968klein,ruffini1969self} and Proca stars (PSs) for vector fields \cite{Brito_2016}. While BSs have been extensively studied since the late 1960s, with many classes of solutions and properties now well understood (see \cite{Schunck_2003,Liebling_2023,Jetzer1992} for reviews), PSs have only been explored more recently, with a growing body of work investigating their structure, dynamics, stability, and phenomenology\footnote{Self-gravitating solitonic configurations supported by massive spin-2 fields have also recently begun to attract attention \cite{Schiappacasse:2025mao,Jain:2021pnk,Aoki:2017ixz}.}; see, e.g., Refs.~\cite{Sanchis-Gual:2017bhw,ignacio_16,Mio:2025fnj,wang_24,Brito_2016,Herdeiro2023:groundstate,Aoki:2022mdn,Martinez:2022wsy,Mio:2026wfh,Lazarte:2025wlw,Zhang:2023rwc,Herdeiro:2020jzx}.

In the present work, we focus on uncharged, nonrotating configurations formed by a massive complex vector (Proca) field without self-interactions. It was recently shown in Ref.~\cite{Herdeiro2023:groundstate} that the ground state of static PSs is axially symmetric, in fact prolate, whereas the spherically symmetric configuration corresponds to an excited state. By contrast, the ground state of static BSs is spherically symmetric. This distinction is especially relevant for EMRIs, since the symmetry (and nature) of the central object may leave direct imprints on the orbital dynamics and emitted radiation.

EMRIs into static, spherically symmetric BSs have been extensively studied in both the Newtonian and relativistic regimes \cite{Macedo_2013,Guo_2019,Annulli:2020l,Destounis:2023khj,Duque_2024,Macedo:2026}. By contrast, EMRIs into PSs have received far less attention. Existing studies connecting EMRIs with Proca fields have considered related but distinct scenarios, involving Proca-haired secondaries \cite{Zi:2024,Zi:2025kerr,Zi:2025qos}, MBHs surrounded by superradiant Proca clouds~\cite{Baumann:2019ztm,Cao:2024wby} or cold vector DM environments around MBHs \cite{Karmakar:2025drp}.
To start addressing this gap, we study EMRIs into static PSs in the Newtonian regime, considering both spherically and axially symmetric configurations, with and without a central ``BH,'' here modeled as a central pointlike mass, as a first step toward a fully relativistic description. In the absence of a central BH, these systems may be viewed as ECOs~\cite{Liebling_2023}; in its presence, they may instead serve as a simple model for the core of DM haloes~\cite{Annulli:2020l,Cardoso:2022nzc}. Particular attention is devoted to the axially symmetric ground-state PSs, which remains unexplored in the EMRI context. Our main goal is to compute, in the Newtonian limit, the energy lost by a point particle on a circular, equatorial orbit around spherically and axially symmetric PS configurations, and to compare it with the same quantity obtained in the spherically symmetric ground-state BS case. We find that the energy lost by the particle is similar in the ground-state configurations, with relative differences of at most $\sim 20\%$, while one to two orders of magnitude differences are found when comparing with the spherically symmetric Proca star, when a central ``parasitic" BH is present (see Fig.~\ref{Fig:All-summed-over-plot}).

The remainder of this paper is organized as follows. In Sec.~\ref{sec:framework}, we outline the framework used in this work, including the action and perturbative setup. In Sec.~\ref{sec:NPS-bkg}, we analyse the spherically and axially symmetric PS backgrounds in the Newtonian regime. In Sec.~\ref{sec:linear-pert}, we obtain the equations needed to study linear perturbations around both configurations, while in Sec.~\ref{sec:energy-fluxes} we present the formalism to compute the Proca energy and Noether charge fluxes generated by a point particle on a circular, equatorial orbit around a Newtonian PS. In Sec.~\ref{sec:methods}, we describe the methods used to solve the perturbative equations and compute the fluxes. Section~\ref{sec:numerical_results} presents the main results of this paper. Specifically, we compute the orbital energy loss rate during adiabatic inspirals of a point particle into Newtonian PSs, considering both spherically and axially symmetric PSs, as well as the impact of including a central BH in the configuration (Sec.~\ref{subsec:w_wo_BH_numerical-results}). We then focus on the case with a central parasitic BH (Sec.~\ref{subsec:with_a_parasitic_BH}) and compare the corresponding net loss rates with those of the spherically symmetric Newtonian BS case in Sec.~\ref{subsec:compare_NBS}. Finally, in Sec.~\ref{sec:conclusions}, we summarize our findings and discuss future prospects. Additional technical details are also given in the appendices. We adopt the metric signature $(-,+,+,+)$ and work in geometrized units with $G=c=1$. Repeated upper/lower Greek indices run over spacetime components, $\mu \in \{0,1,2,3\}$, while Latin indices run over spatial components, $i \in \{1,2,3\}$, unless stated otherwise.

\section{Framework}
\label{sec:framework}

\subsection{Action and field equations}
\label{subsec:action_and_eqs}

We consider a complex Proca field $\mathcal{A}_\mu$ minimally coupled to gravity, described by the action
\begin{equation}
    \mathcal{S}=\int \dd^4 x \sqrt{-g} \left( \frac{R}{16 \pi} -\frac{1}{4} \mathcal{F}_{\mu \nu} \bar{\mathcal{F}}^{\mu \nu} - \frac{1}{2} \mu^2 \mathcal{A}_\mu \bar{\mathcal{A}}^\mu  + \mathcal{L}^m \right)  ,
\label{eq:action-EP}
\end{equation}
where $\mathcal{L}^m$ represents additional matter fields, assumed to be minimally coupled to gravity as well, and an overbar indicates complex conjugation. Here, $\hbar\mu$ denotes the Proca field mass, with $\mathcal{F}_{\mu \nu}=\nabla_\mu \mathcal{A} _\nu - \nabla_\nu \mathcal{A}_\mu$ the field-strength tensor, $R$ the Ricci scalar and $g\equiv \det(g_{\mu\nu})$ the determinant of the spacetime metric $g_{\mu\nu}$. Varying the action (\ref{eq:action-EP}) with respect to the metric and the (massive) vector field, we obtain the Einstein-Proca (EP) field equations 
\begin{align}
        G_{\mu\nu} &= 8 \pi (T^\text{Proca}_{\mu \nu}+T^m_{\mu \nu}) \,, \label{eq:EP-system1} \\
        \nabla_\mu \mathcal{F}^{\mu \nu} &= \mu^2 \mathcal{A}^\nu \,,
        \label{eq:EP-system2}
\end{align}
where $G_{\mu\nu}:=R_{\mu \nu} - g_{\mu \nu} R /2$ is the Einstein tensor, $T^{\text{Proca}}_{\mu \nu}$ is the stress-energy tensor of the Proca field, 
\begin{align}
\begin{split}
    T^\text{Proca}_{\mu \nu} =& -\mathcal{F}_{\sigma(\mu}\bar{\mathcal{F}}^\sigma_{\nu)}-\frac{1}{4} g_{\mu \nu} \mathcal{F}_{\sigma \rho} \bar{\mathcal{F}}^{\sigma \rho} \\
    &+ \mu^2 \left[ \mathcal{A}_{(\mu} \bar{\mathcal{A}}_{\nu)}-\frac{1}{2} g_{\mu \nu} \mathcal{A}_\sigma \bar{\mathcal{A}}^\sigma \right] \,,
\label{eq:stress-energy-tensor}
\end{split}
\end{align}
and $T^{m}_{\mu \nu}$ corresponds to the stress-energy tensor of additional matter fields. The Proca field possesses a global $U(1)$ symmetry, under the transformation $\mathcal{A}_\mu \rightarrow e^{\mathrm{i} \alpha} \mathcal{A}_\mu$, with $\alpha$ constant, leading to a conserved Noether 4-current 
\begin{equation}
    j^\mu = -\frac{\mathrm{i}}{2} \left[ \bar{\mathcal{F}}^{\mu \nu} \mathcal{A}_\nu - \mathcal{F}^{\mu \nu} \bar{\mathcal{A}}_\nu \right]\,. 
\end{equation}

For a nondissipative system, the conserved current implies the existence of a conserved Noether charge $Q$, obtained by integrating the temporal component, $j^t$, on a spacelike hypersurface $\Sigma$:
\begin{equation}
    Q = \int_\Sigma \dd^3x \sqrt{-g} j^t \,.
\label{eq:noether-charge}
\end{equation}

\subsection{Perturbative setup}
\label{subsec:perturbative-setup}

The matter contribution is provided by the EMRI secondary, which, for sufficiently small mass ratios, can be treated as a point particle of mass $m_p$~\cite{Pound:2021qin}, with Lagrangian density
\begin{equation}
    \mathcal{L}^m = m_p \int \dd\tau \frac{1}{\sqrt{-g}} \delta^{(4)} [x^\mu - x_p^\mu(\tau)] \,,
\label{eq:lagrangian-pp}
\end{equation}
where $x_p^\mu(\tau)$ is the particle's worldline, parametrized by its proper time $\tau$. The corresponding stress-energy tensor is given by
\begin{equation}
    T_m^{\mu\nu} =  m_p \int \dd\tau \frac{1}{\sqrt{-g}} \delta^{(4)} [x^\mu - x_p^\mu(\tau)] \dot{x}_p^\mu \dot{x}_p^\nu \,,
    \label{eq:stress-energy-tensor-pp}
\end{equation}
where $\dot{x}_p^\mu\equiv \dd x_p^\mu/\dd \tau$. We first consider a nonrotating PS background, described by the zeroth-order solution $\{g_{\mu\nu}^{(0)}, \mathcal{A}^{(0)}_{\mu} \}$ of the EP system, Eqs.~(\ref{eq:EP-system1}) and~(\ref{eq:EP-system2}), in the absence of additional matter, $T^m_{\mu\nu}=0$. The EMRI secondary is treated as a small perturbative source on this background. Expanding in the mass-ratio $\varepsilon=m_p/M\ll1$, we write, to leading order in $\varepsilon$,
\begin{align}
    \mathcal{A}_\mu &= \mathcal{A}^{(0)}_{\mu} + \varepsilon  \delta\mathcal{A}_{\mu} + \mathcal{O}(\varepsilon^2) \,, \label{eq:pert-metric-proca1} \\
    g_{\mu\nu} &= g_{\mu\nu}^{(0)} + \varepsilon  \delta g_{\mu\nu} + \mathcal{O}(\varepsilon^2) \,.
    \label{eq:pert-metric-proca2}
\end{align}
Here, $\delta\mathcal{A}_{\mu}$ and $\delta g_{\mu\nu}$ denote the linear Proca field and metric perturbations induced by the secondary. Inserting these expansions into the EP system gives the perturbation equations on the PS background. At leading order, the small body is assumed to move on adiabatic, quasicircular geodesics of $g_{\mu\nu}^{(0)}$, while its stress-energy tensor sources the perturbations. In the Newtonian regime, the perturbation equations and the corresponding point-particle source are specified in Secs.~\ref{sec:linear-pert} and~\ref{sec:energy-fluxes}, respectively.

\subsection{Energy and Noether charge fluxes}

In the presence of the secondary, the Proca field is perturbed, radiating both energy and $U(1)$ Noether charge.
Away from the particle's worldline, the radiative stress-energy tensor satisfies $\nabla_\mu \delta T^{\mu\nu}_\Proca=0$. For a stationary, asymptotically flat background with timelike Killing vector $\bm{\xi}_t=-\partial_t$, conservation of the radiative stress-energy tensor implies conservation of the energy current 
$\delta J_E^\mu\equiv g^{\mu \rho} \delta T^{\rm Proca}_{\rho\nu}\xi^\nu$. Gauss's theorem then gives the outgoing energy flux radiated through a $2$-sphere at infinity,
\begin{equation}
    \dot{E}^\text{rad} = - \lim_{r \to \infty} r^2 \int \dd\Omega \,\delta T^\Proca_{rt} \,,
\label{eq:E-rad-Proca}
\end{equation}
where $\dd\Omega=\dd\theta \dd\varphi \sin \theta$. To leading order, $\mathcal{O}(\varepsilon^2)$, and asymptotically, the perturbed Proca stress-energy tensor reads
\begin{align}
\begin{split}
    \delta T^\Proca_{\mu \nu} (r \to \infty) &\sim -\delta\mathcal{F}_{\sigma(\mu}\delta\bar{\mathcal{F}}^\sigma_{\nu)}-\frac{1}{4} \eta_{\mu \nu} \delta\mathcal{F}_{\sigma \rho} \delta\bar{\mathcal{F}}^{\sigma \rho} \\
    &+ \mu^2 \left[ \delta\mathcal{A}_{(\mu} \delta\bar{\mathcal{A}}_{\nu)}-\frac{1}{2} \eta_{\mu \nu} \delta\mathcal{A}_\sigma \delta\bar{\mathcal{A}}^\sigma \right]\,.
\end{split}\label{eq:stress-energy-tensor-pert}
\end{align}

Since the secondary is neutral, the $U(1)$ current remains conserved in the asymptotic region, $\nabla_\mu \delta j^\mu=0$, and the outgoing Noether charge flux is
\begin{equation}
    \dot{Q}^\text{rad} = -\lim_{r\to\infty} r^2 \int \dd\Omega \,\delta j_r \,,
\label{eq:Q-flux-Proca}
\end{equation}
which we interpret as the rate at which massive vector quanta (or particles) are carried away to infinity. The perturbed current $\delta j_\mu$, at leading order $\mathcal{O}(\varepsilon^2)$ and asymptotically, is given by
\begin{equation}
    \delta j_\mu (r\to \infty) \sim - \frac{\mathrm{i}}{2} \eta^{\sigma \rho} \left[ \delta\bar{\mathcal{F}}_{\mu \rho} \,\delta \mathcal{A}_\sigma - \delta \mathcal{F}_{\mu \rho} \,\delta \bar{\mathcal{A}}_\sigma \right] \,. 
\label{eq:noether-charge-pert}
\end{equation}

In the Proca sector, the \textit{total} energy lost contains two contributions: the energy radiated to infinity by the Proca field, and the change in the PS mass associated with the radiated $U(1)$ Noether charge. Following Ref.~\cite{Annulli:2020l}, we may relate the PS mass to the Noether charge in the nonrelativistic limit through $M=\mu Q$, which, at leading order, yields $\dot M=\mu\,\dot Q$. Thus, we define the total energy lost by the perturber due to energy exchange with the Proca field, per unit time, as
\begin{align}
\begin{split}
    \dot{E}^{\text{lost}} &\equiv \dot{E}^{\text{rad}} + \dot{M} \\&=\dot{E}^{\text{rad}} + \mu \dot{Q}^{\text{rad}} \,.
\label{eq:Eloss-Proca}
\end{split}
\end{align}

Including emission from GWs, the total loss rate of the secondary’s orbital energy is
\begin{equation}
  \frac{\dd E}{\dd t} \equiv - \dot{E}^g - \dot{E}^{\text{lost}}\,,
\end{equation}
where $\dot E^{g}$ is the GW energy flux at infinity\footnote{Note that the GW energy flux cannot be computed within our Newtonian setup, since one needs to account for
relativistic effects. However, to leading order, the GW energy flux can be easily computed using the quadrupole formula \cite{Maggiore:2007}.}.

\section{Newtonian Proca stars}
\label{sec:NPS-bkg}

In this section, we study localized, self-gravitating solutions of the EP system (\ref{eq:EP-system1})--(\ref{eq:EP-system2}), without additional fields, known as PSs \cite{Brito_2016}, in a Newtonian regime.
Following the notation of Ref.~\cite{Annulli:2020l}, we adopt a time-periodic ansatz for the Proca field $\mathcal{A}_\mu(x^\mu)$ of the form 
\begin{equation}
    \mathcal{A}_\mu(t, \mathbf{x}) = \mathcal{A}_\mu(\mathbf{x}) e^{-\mathrm{i} \Omega t} \,.
\label{eq:proca-ansatz-bkgg}
\end{equation}

The harmonic frequency, $\Omega=\mu-\gamma$, can be thought of as the energy of the individual vector particles, given by their rest mass $\mu$ and binding energy parameter $\gamma$. In order to obtain the so-called Newtonian Proca stars (NPSs), we use the ``Newtonian'' spacetime metric in Newtonian or harmonic gauge:
\begin{equation}
\dd s^2 \approx -(1 + 2U)\,\dd t^2 + (1-2U)\delta_{ij}\dd x^i \dd x^j \,,
\label{eq:newtonian-metric}
\end{equation}
with a weak gravitational potential $|U(r)| \ll 1$ and a small amplitude vector field $|\mathcal{A}_\mu| \ll 1$. In this limit, the energy of the individual vector particles is approximately given by their rest mass, i.e., $0 < \gamma \ll \mu$. The EP system reduces to a ``vectorial" Schr\"odinger-Poisson (SP) system (see Appendix~\ref{app:SP}) given by
\begin{align}
\mathrm{i} \partial_t \tilde{\mathcal{A}}_j &= - \frac{1}{2 \mu} \nabla^2 \tilde{\mathcal{A}}_j + \mu U \tilde{\mathcal{A}}_j \,, \label{eq:SP-system-ps1}\\ 
\nabla^2 U &= 4\pi\mu \tilde{\mathcal{A}}_{j}(\tilde{\mathcal{A}}^{j})^* \,,
\label{eq:SP-system-ps2}
\end{align}
denoting with an asterisk the respective complex conjugate. Here, the Schr\"{o}dinger field $\tilde{\mathcal{A}}_j$ is related to the Proca field $\mathcal{A}_j$ through $\tilde{\mathcal{A}}_j(x^\mu) \equiv \sqrt{\mu} e^{\mathrm{i}\mu t} \mathcal{A}_j (x^\mu)$. The SP system contains only the spatial components $\mathcal{A}_j$ of the Proca field, since the temporal component $\mathcal{A}_0$ is an auxiliary field fixed by the Lorenz constraint
\begin{equation}
    \nabla_\mu \mathcal{A}^\mu = 0 \,,
    \label{eq:lorenz-const}
\end{equation}
which is not a gauge choice, due to the mass term in the Proca action (Eq.~\eqref{eq:action-EP}, with $\mathcal{L}^m=0$), but a dynamical requirement \cite{Adshead:2021l,Brito_2016}. Accordingly, we henceforth write the Proca ansatz [Eq.~\eqref{eq:proca-ansatz-bkgg}] only in terms of the independent spatial components, $\mathcal{A}_j(x^\mu) = \mathcal{A}_j(\mathbf{x} )e^{-\mathrm{i} \Omega t}$.

\subsection{Background configurations}
\label{subsec:bkg-configs}

We are interested in obtaining nonrotating background NPS configurations. 
We focus on axially symmetric solutions of the Proca field equations on flat spacetime, described in spherical coordinates by the Proca ansatz \cite{Herdeiro2023:groundstate}: 
\begin{equation}
    \mathcal{A}=  \left(  \mathrm{i} H_0(r,\theta)\dd t + \frac{H_1(r,\theta)}{r} \dd r + H_2(r,\theta)\dd\theta  \right) \!  e^{-\mathrm{i} \Omega t} \,,
\label{eq:PS-ansatz-Herdeiro}
\end{equation}
with $H_i(r,\theta)=\sum_\ell c_{i,\ell} h_{i,\ell}(r) Y_{\ell 0}(\theta)$, for $i=0,1$ and $H_2(r,\theta)=\sum_\ell c_{2,\ell} h_{2,\ell}(r) \partial_\theta Y_{\ell 0}(\theta)$. Here, $c_{i,\ell}$, $h_{i,\ell}(r)$ and $Y_{\ell 0}(\theta)$ correspond to constants, radial functions, and $\varphi$-independent spherical harmonics (necessary for the solutions to be nonrotating), respectively, for $i = 0, 1, 2$ and $\ell \in \mathbb{N}_0$.

\subsubsection{Excited (hedgehog) solution}
\label{subsubsec:hedge-bkg}

We start by considering the spherically symmetric ($\ell=0$) solution of NPSs, commonly referred to as \textit{hedgehog} solution \cite{Adshead:2021l}. From the ansatz in Eq.~(\ref{eq:PS-ansatz-Herdeiro}), we write the time-independent \textit{spatial} components of the Proca field in spherical coordinates, $\mathcal{A}_j(\mathbf{x}) \equiv \left( \mathcal{A}_r, \mathcal{A}_\theta, \mathcal{A}_\varphi \right)$, as
\begin{equation}
\begin{split}
    \mathcal{A}_j(\mathbf{x}) &\equiv \Big( c_{1,0} \frac{h_{1,0}(r)}{r} Y_{00}\,,\, c_{2,0} h_{2,0}(r) \partial_\theta Y_{00}\,,\,0 \Big) \\
    & \equiv  \Big(H(r),0,0 \Big) \,,
    \label{eq:ansatz-hedge-herdeiro}
\end{split}
\end{equation}
where we defined $H(r) \equiv c_{1,0} h_{1,0}(r)/(\sqrt{4\pi}r)$. This is the ansatz that was considered in Ref.~\cite{Adshead:2021l}.

Using the spatial ansatz above in the SP system, and recalling the harmonic time dependence $e^{-\mathrm{i} \Omega t}$ of the Proca field [Eq.~\eqref{eq:proca-ansatz-bkgg}], we obtain the following system of radial differential equations:
\begin{align}
\partial_r^2 H + \frac{2}{r}\partial_r H - \frac{2}{r^2} H - 2 \mu (\mu U + \gamma) H &=0 \,,  \label{eq:SP-bkg-hedge1}\\
\partial_r^2 U + \frac{2}{r} \partial_r U - 4 \pi \mu^2 H^2&= 0 \,.
\label{eq:SP-bkg-hedge2}
\end{align}

This system is invariant under the transformation 
\begin{equation}
    (H,U,\gamma) \rightarrow \lambda^2 (\tilde{H},\tilde{U},\tilde{\gamma})\,, \quad r \rightarrow \tilde{r}/\lambda \,,
\end{equation}
resulting in the NPS mass scaling as $\Mnps \rightarrow \lambda \tilde{M}_\text{\tiny NPS}$, where the NPS mass is given by
\begin{equation}
    \Mnps = 4 \pi \mu^2 \int_0^\infty \dd r\,r^2 |H(r)|^2 \,.
\label{eq:NPS-mass-hedge}
\end{equation}

Such scale invariance allows us to ignore the weak-field constraints on $H, \,U, \,\gamma$ when solving the equations, since one can always rescale the solutions so that they are within the regime of validity of the Newtonian approximation~\cite{Annulli:2020l}. Moreover, the particular rescaling $ (H,U,\gamma / \mu) \rightarrow \lambda^2 (\tilde{H},\tilde{U},\tilde{\gamma})$ and $r \mu \rightarrow \tilde{r}/\lambda$ leaves the SP system (\ref{eq:SP-bkg-hedge1})--(\ref{eq:SP-bkg-hedge2}) independent of $\mu$. Following Ref.~\cite{Annulli:2020l}, we define $\lambda \equiv \mu \Mnps$ and write the scale-invariant relations:
\begin{equation}
\begin{aligned}
\tilde{H} &= H / (\Mnps^2 \mu^2)\,, &
\quad \tilde{U} &= U / (\Mnps^2 \mu^2)\,, \\
\tilde{r} &= r \Mnps \mu^2\,, &
\tilde{\gamma} &= \gamma / (\Mnps^2 \mu^3) \,.
\label{eq:proca-rescalling}
\end{aligned}
\end{equation}

\begin{figure}
\centering
\includegraphics[width=0.45\textwidth,clip]{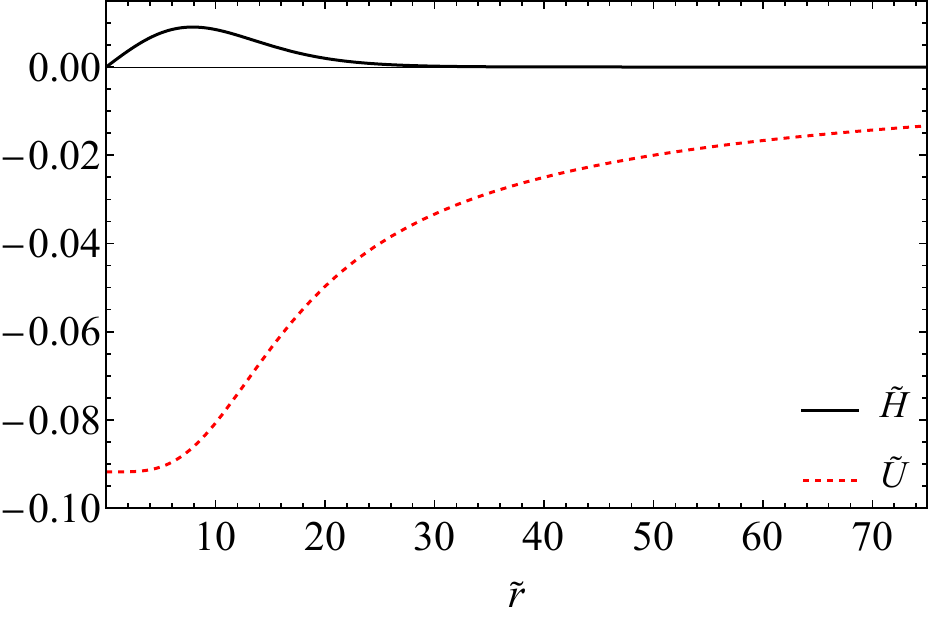}
\caption{Hedgehog solution, given in terms of the rescaled quantities (\ref{eq:proca-rescalling}). 
The rescaled gravitational potential near the origin is $\tilde{U}(0) \simeq -0.0918$. The fields $\tilde{H}$ and $\tilde{U}$ are characterized by the rescaled energy parameter $\tilde{\gamma} \simeq 0.0541072$.}
\label{fig:bkg-hedgehog}       
\end{figure}

Expanding $U(r)$ and $H(r)$ as power series at the origin and demanding asymptotic flatness at infinity, we solve the eigenvalue problem for the lowest energy parameter $\gamma$ using a shooting method. Direct integration yields the numerical results presented in Fig.~\ref{fig:bkg-hedgehog}, which are consistent with the literature\footnote{Figure~\ref{fig:bkg-hedgehog} is consistent with Fig.~1 from Ref.~\cite{Adshead:2021l}, after the rescaling $\tilde{H} \rightarrow \tilde{H} m_{\rm Pl}^3$, $\tilde{U} \rightarrow \tilde{U} m_{\rm Pl}^4$ and $\tilde{r} \rightarrow \tilde{r} / m_{\rm Pl}^2$. Here, the Planck mass is given by $m_{\rm Pl} = 1 / \sqrt{8\pi}$.}.

Near the center, $U(r)$ is approximately constant, whereas $H(r) \sim r$. For large radii, the gravitational potential
falls off as $U \sim -\Mnps/r$, while the hedgehog radial profile decays according to a Yukawa potential $H \sim e^{-\sqrt{2\mu\gamma}r}/r$.
Since the lowest-energy solution of $H(r)$ has a node at the origin, i.e., $H(0)=0$, the hedgehog solution does not represent the ground state of NPSs. Instead, the spherically symmetric configuration corresponds to an excited state of an NPS~\cite{Herdeiro2023:groundstate}.


\subsubsection{Ground-state solution}
\label{subsubsec:bkg-gs}

Recently, it has been shown that the \textit{true} ground state of static PSs corresponds to a nonspherical configuration with dipolar ($\ell = 1$) symmetry \cite{Herdeiro2023:groundstate}. This solution is axially symmetric and the relativistic solution exhibits a distinctively \textit{prolate} structure\footnote{This prolate structure seems to be a purely relativistic effect, as recently noticed~\cite{Diez-Tejedor:2026fnc}. As we show below, the energy density of the NPS ground state is instead spherically symmetric.}. We will now construct the Newtonian counterpart of the solutions obtained in Ref.~\cite{Herdeiro2023:groundstate}. 

From the Proca ansatz [Eq.~\eqref{eq:PS-ansatz-Herdeiro}], the spatial components are given by
\begin{equation}
    \mathcal{A}_j (\mathbf{x})= \Big( h_1(r)Y_{10}(\theta), h_2(r) \partial_\theta Y_{10}(\theta),0 \Big) \,,
\label{eq:gs-ansatz-1}
\end{equation}
where $h_1(r) = c_{1,1} h_{1,1}(r)/r$, $h_2(r) = c_{2,1} h_{2,1}(r)$ and $Y_{10}(\theta)=\sqrt{3/4 \pi} \cos(\theta)$. Substituting this ansatz into the SP system (\ref{eq:SP-system-ps1})--(\ref{eq:SP-system-ps2}), we find that the radial functions satisfy the relation $h_2(r) = r h_1(r)$, which allows us to rewrite the previous ansatz as
\begin{equation}
    \mathcal{A}_j(\mathbf{x})= \sqrt{\frac{3}{4\pi}} h_1(r) \big(\cos(\theta), -r \sin(\theta),0 \big) \,.
\label{eq:gs-ansatz-2}    
\end{equation}

Up to an overall normalization factor, this ansatz is equivalent to the one considered in Ref.~\cite{Adshead:2021l}, where the following spatial Proca field ansatz in Cartesian coordinates was considered:
\begin{equation}
    \mathcal{A}_j (\mathbf{x}) \equiv \left( \mathcal{A}_x,\mathcal{A}_y,\mathcal{A}_z \right) = h(r) \left( v_x , v_y , v_z \right) \,,
\label{eq:bkg-gs-ansatz-cart}
\end{equation} 
with $v_j$ a constant unit vector $||v_j||=1$. To show the equivalence between Eqs.~\eqref{eq:gs-ansatz-2} and~\eqref{eq:bkg-gs-ansatz-cart}, consider an axially symmetric configuration with $v_j$ pointing along the $z$-axis, i.e., $v_j = (0, 0, 1)$. In spherical coordinates, this corresponds to $v_j = (\cos(\theta), -r \sin(\theta), 0)$ and the ansatz in Eq.~\eqref{eq:bkg-gs-ansatz-cart} becomes
\begin{equation}
    \mathcal{A}_j (\mathbf{x})  \equiv \left( \mathcal{A}_r,\mathcal{A}_\theta,\mathcal{A}_\varphi \right) = h(r) \left( \cos (\theta), -r \sin (\theta), 0 \right) \,.
\label{eq:gs-ansatz-1-our}
\end{equation}
Renaming the radial function $h(r)\equiv h_1(r)$ and multiplying by a factor of $\sqrt{3/4\pi}$, we recover the ansatz previously given in Eq.~\eqref{eq:gs-ansatz-2}. We conclude that the two approaches of Refs.~\cite{Adshead:2021l,Herdeiro2023:groundstate} are equivalent.

\begin{figure}[h]
\centering
\includegraphics[width=0.45\textwidth,clip]{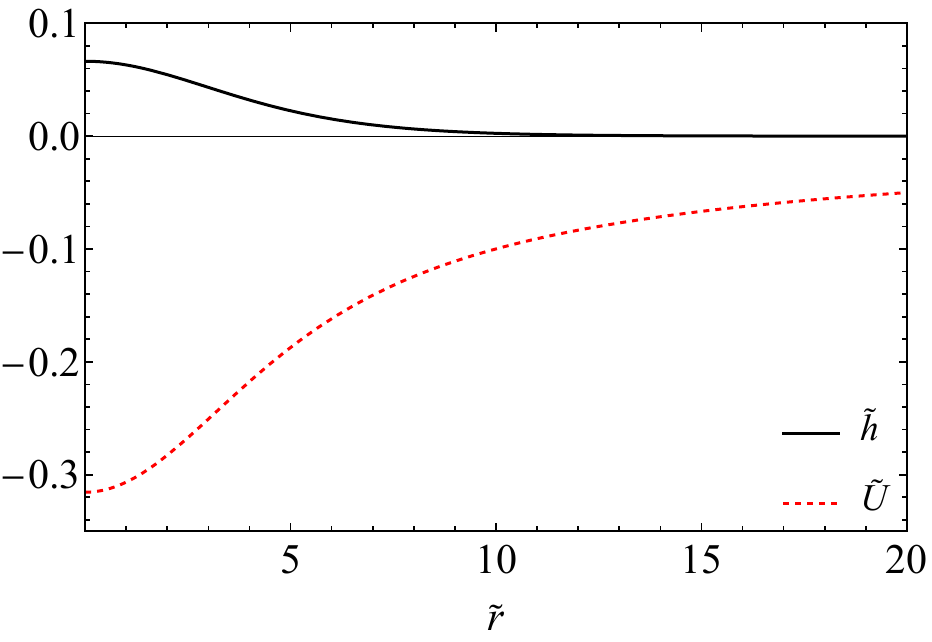}
\caption{Ground-state solution, given in terms of the rescaled quantities (\ref{eq:proca-rescalling}), now for $h(r)$. Near the origin, the rescaled gravitational potential and vector radial profile are $\tilde{U}(0) \simeq -0.31551$ and $\tilde{h}(0) \simeq 0.06633$, respectively. The fields $\tilde{h}$ and $\tilde{U}$  are characterized by the energy parameter $\tilde{\gamma} \simeq 0.162769$.}
\label{fig:NPS-plot}       
\end{figure}

Using the spatial ansatz of Eq.~(\ref{eq:gs-ansatz-1-our}) in the SP system, we now obtain the following radial differential equations:
\begin{align}
\begin{split}
\partial^2_r h + \frac{2}{r} \partial_r h - 2 \mu (\mu U + \gamma)h  &= 0\,, \\ 
\partial^2_r U + \frac{2}{r} \partial_r U - 4 \pi \mu^2 h^2 &= 0\,.
\label{eq:SP-bkg-gs}
\end{split}
\end{align}
This system possesses the same scale-invariance relations as in the hedgehog case [Eq.~\eqref{eq:proca-rescalling}]. As before, we solve the eigenvalue problem for the lowest energy parameter $\gamma$ using a shooting method, imposing regularity at the origin and asymptotic flatness at infinity. The numerical solutions are shown in Fig.~\ref{fig:NPS-plot}, again, consistent with the literature\footnote{Fig.~\ref{fig:NPS-plot} is consistent with Fig.~3 from Ref.~\cite{Adshead:2021l}, after the rescaling $\tilde{h} \rightarrow \tilde{h} \,m_{\rm Pl}^3$, $\tilde{U} \rightarrow \tilde{U} m_{\rm Pl}^4$ and $\tilde{r} \rightarrow \tilde{r} / m_{\rm Pl}^2$.}.

Near the center, both fields $U$ and $h$ are approximately constant. For large distances, the gravitational potential and the vector radial profile fall off as in Sec.~\ref{subsubsec:hedge-bkg}, i.e., $U \sim - \Mnps/r$ and $h \sim e^{-\sqrt{2 \mu \gamma}r}/r$, with the ground-state NPS mass given by Eq.~(\ref{eq:NPS-mass-hedge}), but now integrating $h(r)$. Since the lowest-energy solution of $h(r)$ has no nodes, we may conclude that the axially symmetric configuration of Eq.~(\ref{eq:gs-ansatz-1-our}) indeed represents the ground state of NPSs.

We end this section with some remarks. The radial system of differential equations describing the NPS ground state, Eq.~\eqref{eq:SP-bkg-gs}, is equivalent to the one obtained for the ground state of Newtonian boson stars (NBSs)\footnote{Henceforth, unless stated otherwise, NBSs refer to their ground-state configurations.} (cf. Eq.~(39), Ref.~\cite{Annulli:2020l}). Consequently, unperturbed ground-state solutions for NPSs are indistinguishable from those of NBSs: both have identical radial profiles, as well as the same rescaled energy parameter $\tilde{\gamma}$ (cf. Fig.~2, Ref.~\cite{Annulli:2020l}).  Furthermore, although the ground-state Proca field is not spherically symmetric (Eq.~\ref{eq:gs-ansatz-1-our}), its energy density is spherically symmetric, as can be seen from the Poisson equation, $\nabla^2U(r)=4 \pi \mu^2 h(r)^2$, which depends only on $r$. Therefore, the energy density profile of the ground state of NPSs is identical to the one from NBSs, hence they source the same gravitational potential\footnote{Note that this is only true at the Newtonian level. When considering fully relativistic solutions, the spacetime of the PS ground state is not spherically symmetric, unlike the one of the BS ground state~\cite{Herdeiro2023:groundstate}.}.
We can conclude that the NPSs ground state presents the same scale-invariant mass-radius relation as for NBSs~\cite{Annulli:2020l}
\begin{equation}
    \frac{M_\text{\tiny NPS}}{M_\odot} = 8 \times 10^9 \left( \frac{1 \,\text{kpc}}{R} \right) \left( \frac{10^{-22} \,\text{eV}}{\mu} \right)^2 \,,
\end{equation}
with the radius $R$ corresponding to $98 \%$ of the bosonic star mass.

To conclude this section, we note that the relation $M_\text{\tiny NPS} = \mu \,Q_\text{\tiny NPS}$ allows us to define the number of (vector) particles in the NPS as
\begin{equation}
    Q_\text{\tiny NPS} = 4 \pi \mu \int_0^\infty \dd r\, r^2 |g(r)|^2 \,,
\label{eq:NPS-charge-gs}
\end{equation}
where $g(r)$ denotes the radial profile of either the hedgehog solution, $H(r)$, or the ground-state solution, $h(r)$.

Quite interestingly, both the hedgehog and the ground-state solutions are stable against small perturbations in the Newtonian regime~\cite{Nambo:2025lnu}, unlike what happens in the relativistic regime, where spherically symmetric PS configurations (the relativistic version of the hedgehog configuration~\cite{Diez-Tejedor:2026fnc}) are unstable against nonspherical perturbations~\cite{Herdeiro2023:groundstate}, although the instability timescale becomes arbitrarily large in the Newtonian limit~\cite{Diez-Tejedor:2026fnc}. This suggests that, for dilute Proca stars, both solutions could play a role in astrophysics.

\section{Linear perturbation theory}

\subsection{Sourceless perturbations}
\label{sec:linear-pert}

Having constructed the stationary Newtonian PS backgrounds in the previous section, we now study linear perturbations to these configurations. No additional matter source is included at this stage; the point-particle source associated with the EMRI secondary will be introduced in Sec.~\ref{sec:energy-fluxes}. In the Newtonian limit, the metric perturbation $\delta g_{\mu\nu}$ is encoded in the perturbation of the gravitational potential, $\delta U$. We therefore expand the Proca and gravitational potential fields as
\begin{align}
\begin{split}
    \mathcal{A}_j &= \mathcal{A}_{0j} + \varepsilon  \delta \mathcal{A}_j + \mathcal{O}(\varepsilon^2) \,,\\
    U &= U_0 + \varepsilon \delta U + \mathcal{O}(\varepsilon^2) \,, 
\end{split}\label{eq:ansatz-proca-lin-pert}
\end{align}
where $\varepsilon \ll 1$ is a small parameter controlling the perturbative expansion. We will later identify $\varepsilon$ with the EMRI mass ratio. The fields $\{ \mathcal{A}_{0j}, U_0 \}$ denote the background solutions discussed in the previous section, while $\{ \delta \mathcal{A}_j, \delta U \}$ denote the corresponding linear perturbations. Substituting Eq.~\eqref{eq:ansatz-proca-lin-pert} into the source-free SP system, Eqs.~\eqref{eq:SP-system-ps1} and \eqref{eq:SP-system-ps2}, the zeroth-order terms reproduce the background equations, while at first order in $\varepsilon$, we obtain the perturbed SP system
\begin{align}
    \mathrm{i} \partial_t \delta \mathcal{A}_j &= -\frac{1}{2\mu}\nabla^2 \delta \mathcal{A}_j + (\mu U_0 + \gamma) \delta \mathcal{A}_j + \mu  \mathcal{A}_{0j} \delta U \,, \label{eq:pert-SP-system-NPS-1} \\
     \nabla^2\delta U &= 4 \pi  \left[ \mu^2 \mathcal{A}_{0j} (\delta \mathcal{A}^j + \delta \bar{\mathcal{A}}^j) \right] .
\label{eq:pert-SP-system-NPS-2}
\end{align}

We work in the frequency domain and decompose the Proca field and gravitational potential perturbations into a vector and scalar spherical harmonic basis, respectively:
\begin{align}
    &\delta \mathcal{A}_j = \sum_{\ell,m} \int \frac{\dd\omega}{\sqrt{2\pi}} \Big(\delta \mathbf{A}_{(1)j}^{\omega \ell m} e^{-\mathrm{i} \omega t} +  (\delta \mathbf{A}_{(2)j}^{\omega \ell m})^* e^{\mathrm{i} \omega t} \Big) e^{-\mathrm{i}\Omega t}\,,
\label{eq:fourier-pert-VSH} \\
    &\delta U = \sum_{\ell,m} \int \frac{\dd\omega}{\sqrt{2\pi}r} \left[u^{\omega \ell m} Y_\ell^m e^{-\mathrm{i}\omega t}+(u^{\omega \ell m})^* (Y_\ell^m)^* e^{\mathrm{i}\omega t}\right]\,, \label{eq:ansatz-U-pert}
\end{align}
where $Y^{\ell m} \equiv Y^{\ell m}(\theta,\varphi)$ denotes the scalar spherical harmonics and $\delta\mathbf{A}_{(k)j}^{\omega \ell m}$ ($k=1,2$) take the form~\cite{Rosa_2012,Pani:2012,Zi:2024,Ferrari:2020}
\begin{equation}
    \delta \mathbf{A}_{(k)j}^{\omega\ell m}=
\begin{bmatrix}
0 \\
Z^{\omega\ell m}_{k\mathcal{A}} \,\mathbf{S}^{\ell m}_{a}
\end{bmatrix}
+
\begin{bmatrix}
Z^{\omega\ell m}_k/r \,Y^{\ell m} \\
Z^{\omega\ell m}_{k\mathcal{P}}  \,\mathbf{Y}^{\ell m}_{a}
\end{bmatrix} \,,
\label{eq:VSH-pert-1}
\end{equation}
with
\begin{equation}
\begin{split}
    &\mathbf{Y}_a^{\ell m}= \left( \partial_\theta Y^{\ell m}, \partial_\varphi Y^{\ell m} \right) \,, \\
    &\mathbf{S}_a^{\ell m}= \left( \frac{1}{\sin \theta}\partial_\varphi Y^{\ell m}, - \sin \theta \,\partial_\theta Y^{\ell m} \right)\,.
\end{split}
\end{equation}
Here, $\mathbf{Y}_a^{\ell m}/\mathbf{S}_a^{\ell m}$ denote the polar/axial vector spherical harmonics (VSHs), $\partial_\theta$ and $\partial_\varphi$ denote partial derivatives with respect to $\theta$ and $\varphi$, respectively, while $u^{\omega \ell m}$, $Z^{\omega\ell m}_{k\mathcal{A}}$, $Z_k^{\omega\ell m}$, $Z^{\omega\ell m}_{k\mathcal{P}}$ are complex radial functions. In what follows, we omit the labels $\omega, \ell, m$ on these functions for simplicity, unless needed for clarity. We note that the additional factor $e^{-\mathrm{i}\Omega t}$ in Eq.~\eqref{eq:fourier-pert-VSH} follows from the harmonic time dependence of the Proca ansatz, Eq.~\eqref{eq:proca-ansatz-bkgg}. Furthermore, the complex conjugate terms are added in Eq.~\eqref{eq:ansatz-U-pert} to ensure that $\delta U$ is real-valued in full generality.

\subsubsection{Hedgehog configuration}
\label{subsubsec:hedge-small}

Substituting Eqs.~\eqref{eq:fourier-pert-VSH} and \eqref{eq:ansatz-U-pert} into Eqs.~\eqref{eq:pert-SP-system-NPS-1} and \eqref{eq:pert-SP-system-NPS-2}, and using the derivative identities and orthogonality relations of the spherical harmonics, yields the following system of equations for perturbations of the hedgehog configuration:
\begin{align}
    &\hat{\mathcal{D}}_0 u = 4 \pi  \mu ^2 H_0 \left( Z_1 + Z_2 \right) \,, 
     \label{eq:master-hedge-poisson} \\
     &\hat{\mathcal{D}}_k Z_{k} = 2 \mu^2 H_0 u + \frac{2}{r^2}Z_k - \frac{2\ell (\ell+1)}{r^2} Z_{k \mathcal{P}} \,,
     \label{eq:master-hegde-proca1} \\
     &\hat{\mathcal{D}}_k Z_{k \mathcal{P}} = - \frac{2}{r^2} Z_k \,,
     \label{eq:master-hegde-proca2} \\
     &\hat{\mathcal{D}}_k Z_{k \mathcal{A}} = 0\,,
     \label{eq:master-hegde-proca3}
\end{align}
where we defined the differential operators
\begin{equation}
    \begin{split}
        \hat{\mathcal{D}}_0 &\equiv \frac{\dd^2}{\dd r^2} - \frac{\ell(\ell+1)}{r^2} \,, \\
          \hat{\mathcal{D}}_k &\equiv  \hat{\mathcal{D}}_0 - 2\mu(\gamma - \mathrm{sgn}(\omega_k) \omega+\mu U_0) \,, 
    \end{split}
\end{equation}
for $k=1,2$, with $\mathrm{sgn}(\omega_1)=+1$ and $\mathrm{sgn}(\omega_2)=-1$. 

Following Ref.~\cite{Lopes:2024}, it proves useful to decouple the differential equations for the polar variables $\{ Z_k, Z_{k \mathcal{P}}  \}$, Eqs.~\eqref{eq:master-hegde-proca1} and~\eqref{eq:master-hegde-proca2}, by introducing the linear transformation
\begin{equation}
\begin{aligned}
    Z_k(r) &= \ell \,q_{a_k}(r) - (\ell+1) \,q_{b_k}(r)\,, \\
    Z_{k\mathcal{P}}(r) &= q_{a_k}(r) + q_{b_k}(r) \,,
\label{eq:transformation-qa-qb}
\end{aligned}
\end{equation}
for $k=1,2$. This yields the decoupled hedgehog master equations:
\begin{align}
    &\hat{\mathcal{D}}_0 u = 4 \pi  \mu ^2 H_0 \Big[ \ell \left( q_{a_1} + q_{a_2} \right) - (\ell + 1) 
 \left( q_{b_1} + q_{b_2} \right) \Big] \,,
     \label{eq:master-hedge-poisson-qa-qb} \\
    &\hat{\mathcal{D}}_k q_{a_k} = - \frac{2 \ell}{r^2}   q_{a_k} +  \frac{2 \mu^2 H_0}{2 \ell + 1}u \,,
    \label{eq:master-hedge-qa} \\
    &\hat{\mathcal{D}}_k q_{b_k} =  \frac{2 (\ell+1)}{r^2}  q_{b_k} - \frac{2 \mu^2 H_0 }{2 \ell + 1} u \,,
    \label{eq:master-hedge-qb} \\
    &\hat{\mathcal{D}}_k Z_{k \mathcal{A}} = 0 \,.
     \label{eq:master-hegde-proca3-axial}
\end{align}
We note that $ q_{a_k}$ and $q_{b_k}$ are still coupled through Eq.~\eqref{eq:master-hedge-poisson-qa-qb}. However, as we will discuss later, this set of master equations proves useful in approximations where one ``freezes'' gravitational perturbations. Also, notice that the axial sector completely decouples from the polar sector and that different $\ell,m$ modes also decouple, due to the spherical symmetry of the hedgehog configuration.

\subsubsection{Ground-state configuration}
\label{subsec:gs-NPS-small}

Proceeding analogously to the hedgehog case, 
but now also using the identities~\cite{PhysRevD.Kojima}:
\begin{align}
    \cos \theta Y^\ell &= \mathcal{Q}_{\ell+1} Y^{\ell+1} + \mathcal{Q}_\ell Y^{\ell-1}\,, \label{eq:id1} \\ 
    \sin \theta \, \partial_\theta Y^\ell &= \mathcal{Q}_{\ell+1} \ell Y^{\ell+1} - \mathcal{Q}_\ell (\ell+1) Y^{\ell-1}\,, \label{eq:id2} \\
    \cos^2 \theta \, Y^\ell &= \left(\mathcal{Q}_{\ell+1}^2 + \mathcal{Q}_\ell^2\right) Y^\ell \notag \\
    &\quad + \mathcal{Q}_{\ell+1} \mathcal{Q}_{\ell+2} Y^{\ell+2} + \mathcal{Q}_\ell \mathcal{Q}_{\ell-1} Y^{\ell-2}\,, \label{eq:id3}
\end{align}
we obtain the following set of equations for linear perturbations of the ground-state configuration:
\begin{align}
\hat{\mathcal{D}}_0 u^\ell &= 4 \pi  \mu ^2 h_0 \left[ \mathcal{Q}_\ell \alpha_{\ell-1} + \mathcal{Q}_{\ell+1}\gamma_{\ell+1} - \mathrm{i}\,m \beta_\ell  \right], \label{eq:master-poisson}
\\
\hat{\mathcal{D}}_k Z_{k}^\ell  &= 2 \mu ^2 h_0 \left(\mathcal{Q}_\ell u^{\ell-1} + \mathcal{Q}_{\ell+1} u^{\ell+1}\right) + \Lambda_\ell \,, \label{eq:master-proca1}
\\
\mathrm{i}\,m \hat{\mathcal{D}}_k Z_{k\mathcal{P}}^\ell
&= -(\ell-1)\mathcal{Q}_\ell \tilde{\alpha}_{\ell-1}
 + (\ell+2)\mathcal{Q}_{\ell+1}\tilde{\alpha}_{\ell+1}
 + \mathrm{i}\,m \tilde{\beta}_\ell \,,
\label{eq:master-proca2}
\\
\mathrm{i}\,m  \hat{\mathcal{D}}_k Z_{k\mathcal{A}}^\ell 
&= (\ell-1)\mathcal{Q}_\ell \lambda_{\ell-1} - (\ell+2)\mathcal{Q}_{\ell+1} \lambda_{\ell+1} \notag\\
&\quad+2 \mu ^2 h_0 \Big[\mathcal{Q}_{\ell-1} \mathcal{Q}_\ell u^{\ell-2} + \mathcal{Q}_{\ell+1} \mathcal{Q}_{\ell+2} u^{\ell+2} \notag\\
&\quad \qquad \qquad \qquad + \left(\mathcal{Q}_\ell^2+\mathcal{Q}_{\ell+1}^2-1\right) u^\ell\Big]\,, \label{eq:master-proca3}
\end{align}
where the coefficient $\mathcal{Q}_\ell = \sqrt{ \frac{\ell^2-m^2}{4\ell^2-1}}$ is related to the Clebsch-Gordan coefficients. The functions $\alpha_\ell,\gamma_\ell,\Lambda_\ell,\tilde{\beta}_\ell,\lambda_\ell$ are linear combinations of the polar perturbations and their derivatives, whereas $\beta_\ell$ and $\tilde{\alpha}_\ell$ are linear combinations of the axial perturbations and their derivatives:
\begin{equation}
\begin{split}
    \alpha_{\ell} &= -\ell \left( Z_{1\mathcal{P}}^\ell + Z_{2\mathcal{P}}^\ell \right) + \left( Z_1^\ell + Z_2^\ell \right) \,,\\
    \gamma_{\ell} & = (\ell+1) \left( Z_{1\mathcal{P}}^\ell + Z_{2\mathcal{P}}^\ell \right) + \left( Z_1^\ell + Z_2^\ell \right) \,,\\
    \beta_\ell &= Z_{1\mathcal{A}}^\ell+Z_{2\mathcal{A}}^\ell \,,\\
    \Lambda_\ell &= -\frac{2 \ell (\ell+1) Z_{k\mathcal{P}}^\ell}{r^2} + \frac{2}{r^2}  Z_k^\ell \,,\\
    \tilde{\alpha}_{\ell} &= -\hat{\mathcal{D}}_k Z_{k\mathcal{A}}^\ell \,,\\
    \tilde{\beta}_\ell &= -\frac{2}{r^2} Z_k^\ell \,,\\
    \lambda_{\ell} &=  - \frac{2 Z_k^\ell}{r^2} - \hat{\mathcal{D}}_k Z_{k\mathcal{P}}^\ell \,.
\end{split}
\end{equation}

Unlike for the hedgehog configuration, the resulting system now consists of coupled second-order ordinary differential equations, in which axial and polar perturbations mix across different multipoles $\ell$, due to the fact that the ground-state PS is not spherically symmetric.

Finally, we notice that both the hedgehog and ground-state master equations satisfy the symmetry relation $\mathcal{Z}_1(\omega,\ell,m;r) = (-1)^m \mathcal{Z}_2(-\omega,\ell,-m;r)^*$, where we have defined $\mathcal{Z}_k\equiv \{Z_k,Z_{k\mathcal{P}},Z_{k\mathcal{A}} \}$.

\subsection{Point particle on a circular, equatorial orbit}
\label{sec:energy-fluxes} 

We now include the point-particle source introduced in Sec.~\ref{subsec:perturbative-setup}, described by the Lagrangian density~\eqref{eq:lagrangian-pp} and stress-energy tensor~\eqref{eq:stress-energy-tensor-pp}. In the Newtonian limit, the external perturber contributes to the Poisson equation only through its energy density~\cite{Annulli:2020l}. Thus, relative to the source-free SP perturbation equations of Sec.~\ref{sec:linear-pert}, Eq.~\eqref{eq:pert-SP-system-NPS-1} remains unchanged, while Eq.~\eqref{eq:pert-SP-system-NPS-2} becomes
\begin{equation}
    \nabla^2\delta U = 4 \pi  \left[ \mu^2 \mathcal{A}_{0j} (\delta \mathcal{A}^j + \delta \bar{\mathcal{A}}^j) + P\right]\,,
    \label{eq:SP-pert-poisson-source}
\end{equation}
where $P$ is the Newtonian mass density of the localized perturber
\begin{equation}
    P \equiv m_p\frac{\delta(r-r_p(t))}{r^2}\frac{\delta(\theta-\theta_p(t))}{\sin \theta} \delta (\varphi - \varphi_p(t))\,.
\end{equation}

Following the previous section, we decompose the external source in the same frequency domain and spherical harmonic basis used for the Proca and gravitational potential perturbations\footnote{The source $P$ is real-valued, and its Fourier components obey the symmetry condition $p(\omega,\ell,m;r) = (-1)^m p(-\omega,\ell,-m;r)^*$. As before, we suppress the mode labels $\omega, \ell, m$ on the function $p^{\omega \ell m}(r)$ when no ambiguity arises.},
\begin{equation}
    P = \sum_{\ell,m} \int \frac{\dd\omega}{\sqrt{2\pi}r} \Big( p^{\omega \ell m} Y_\ell^m e^{-\mathrm{i}\omega t}+(p^{\omega \ell m})^* (Y_\ell^m)^* e^{\mathrm{i}\omega t} \Big)\,,
    \label{eq:fourier-pert-P}
\end{equation}
where $p^{\omega \ell m}(r)$ is a complex radial function given by
\begin{equation}
    p(r) \equiv \frac{r}{2\sqrt{2\pi}} \int \dd t \dd\Omega \,P \,(Y^{\ell m})^* e^{\mathrm{i} \omega t}\,.
\label{eq:point-like-fourier}
\end{equation}
With this decomposition, the source term $P$ in Eq.~\eqref{eq:SP-pert-poisson-source} contributes an additional term $4 \pi p(r)$ to the right-hand side of the Poisson master equations derived in Sec.~\ref{sec:linear-pert}.

For a point particle on a circular, equatorial orbit,
\begin{equation}
    r_p(t)=r_{\rm orb}\,,
    \qquad
    \theta_p(t)=\frac{\pi}{2}\,,
    \qquad
    \varphi_p(t)=\omega_{\rm orb}t\,,
\end{equation}
and the radial source coefficients reduce to
\begin{equation}
    p = m_p \sqrt{\frac{\pi}{2}} \frac{Y^{\ell m} (\pi/2,0)}{r} \delta(r-r_{\rm orb}) \delta (\omega-m\omega_{\rm orb}) \,.
\label{eq:source-p}
\end{equation}

\subsubsection{Radiated energy and charge fluxes}
\label{subsec:energy-charge-fluxes}

Using the symmetry relation between $\mathcal{Z}_1$ and $\mathcal{Z}_2$, discussed at the end of Sec.~\ref{sec:linear-pert}, one can eliminate one of the two sets of perturbation functions. We choose to eliminate $\mathcal{Z}_2$ without loss of generality. The radiated energy~\eqref{eq:E-rad-Proca} and Noether charge fluxes~\eqref{eq:Q-flux-Proca} are computed from the perturbed Proca stress-energy tensor~\eqref{eq:stress-energy-tensor-pert} and Noether current~\eqref{eq:noether-charge-pert}. Using the orthogonality relations of the VSHs, one obtains the final expressions for $\dot{E}^\text{rad}$ and $\dot{Q}^\text{rad}$, given by (see Appendix~\ref{app:Erad-Qrad-flux})
\begin{align}
    &\dot{E}^\text{rad} = \frac{2}{\pi} \sum_{\ell,m} \Re \left[ \sqrt{2\mu(m \omega_{\rm orb} - \gamma)} \right]   (\mu-\gamma+m\omega_{\rm orb}) \notag\\
     &  \times \Bigg[ \ell (\ell+1) \left(|Z_{1\mathcal{P}}^\infty|^2+|Z_{1\mathcal{A}}^\infty|^2\right)  + \frac{|Z_1^\infty|^2 \mu^2}{(\mu-\gamma+m\omega_{\rm orb})^2} \Bigg]\,, \label{eq:Erad-expression2}\\
     &\dot{Q}^\text{rad} = -\frac{2}{\pi} \sum_{\ell,m} \Re \left[ \sqrt{2\mu(m \omega_{\rm orb} - \gamma)} \right] \times \notag\\
     &  \Bigg[  \ell (\ell+1) \left(|Z_{1\mathcal{P}}^\infty|^2+|Z_{1\mathcal{A}}^\infty|^2\right) + \frac{|Z_1^\infty|^2 \mu^2}{(\mu-\gamma+m\omega_{\rm orb})^2} \Bigg]\,. \label{eq:Qrad-expression2}
\end{align}

As can be seen, the radiated energy and charge fluxes satisfy $\dot{E}^\text{rad} =  - \dot{Q}^\text{rad} (\mu-\gamma + m \omega_{\rm orb})$\footnote{The same relation holds for NBSs; see Ref.~\cite{Annulli:2020l}.}. Combining this relation with Eq.~\eqref{eq:Eloss-Proca}, the total energy lost (per unit time) by the perturber into the Proca sector is given by
\begin{equation}
\begin{split}
     &\dot{E}^\text{lost} = \frac{2}{\pi} \sum_{\ell,m} \Re \left[ \sqrt{2\mu(m \omega_{\rm orb} - \gamma)} \right]   (m\omega_{\rm orb}-\gamma) \,\times\\
     &   \Bigg[ \ell (\ell+1) \left(|Z_{1\mathcal{P}}^\infty|^2+|Z_{1\mathcal{A}}^\infty|^2\right)  + \frac{|Z_1^\infty|^2 \mu^2}{(\mu-\gamma+m\omega_{\rm orb})^2} \Bigg] \,.
\end{split} \label{eq:Eloss-expression2}
\end{equation}
Notice that the square-root factor restricts the fluxes to modes with $m > 0$. Since $\ell \geq |m|$, only multipoles with $\ell > 0$ contribute.

\section{Methods}
\label{sec:methods}

In this section, we extend the framework developed in Sec.~\ref{sec:linear-pert} by including the point-particle source, as described in Sec.~\ref{sec:energy-fluxes}, and discuss the methods we employ to compute the radiated fluxes.

\subsection{Hedgehog configuration}
\label{subsec:NPS-hedge}
As discussed in the previous section, the external perturber enters only through the Poisson equation. For the hedgehog configuration, the Poisson master equation~\eqref{eq:master-hedge-poisson-qa-qb} is modified to
\begin{equation}
\begin{aligned}
     \hat{\mathcal{D}}_0 u   
     = 4 \pi &\Big[ \mu ^2 H_0  \,\ell \left( q_{a_1} + q_{a_2} \right) \\
     &- \mu ^2 H_0 (\ell + 1) 
 \left( q_{b_1} + q_{b_2} \right) + p \Big] \,,
 \label{eq:poisson-Z-qa-qb}
\end{aligned}
 \end{equation}
leading, together with Eqs.~\eqref{eq:master-hedge-qa} and \eqref{eq:master-hedge-qb}, to an inhomogeneous system of ordinary differential equations (ODEs). Notice that, since the axial perturbation equation~\eqref{eq:master-hegde-proca3-axial} is decoupled from this system, the axial sector is not excited by the point particle for the hedgehog configuration, and we can therefore set $Z_{k\mathcal{A}}=0$.

The polar sector system of ODEs can then be written in matrix form as
\begin{equation}
    \partial_r \mathbf{X} - V_B(r) \mathbf{X} = \mathbf{P}\,,
\label{eq:inhomogeneous-Matrix-scalar}
\end{equation}
with source vector $\mathbf{P}$ and perturbation vector $\mathbf{X}$ given by 
\begin{equation}
\begin{aligned}
    &\mathbf{P}\equiv(0,0,0,0,0,0,0,0,0,4\pi p)^T\,,\\
    &\mathbf{X} \equiv \left( q_{a_1}, q_{a_2}, q_{b_1}, q_{b_2}, u, \partial_r q_{a_1}, \partial_r q_{a_2}, \partial_r q_{b_1}, \partial_r q_{b_2}, \partial_r u \right)^T\,,
\end{aligned}
\end{equation}
and matrix $V_B$ written in block form as
\begin{equation}
\scalebox{1.0}{$
V_B(r)=
\begin{pmatrix}
    \mathbf{0}_{5 \times 5} & \quad  \mathds{1}_5  \\
    \\
    \tilde{V}_B(r) & \quad \mathbf{0}_{5 \times 5}
\end{pmatrix}$}\,,
\label{eq:block-form}
\end{equation}
where $\tilde{V}_B$ is given by
\begin{equation}
\scalebox{0.67}{$
\begin{pmatrix}
 2\mu(\gamma - \omega) + V_a & 0 & 0 & 0 & \dfrac{2 \mu^2 H_0}{2\ell + 1} \\
 0 & 2\mu(\gamma + \omega) + V_a & 0 & 0 & \dfrac{2 \mu^2 H_0}{2\ell + 1} \\
 0 & 0 & 2\mu(\gamma - \omega) + V_b & 0 & -\dfrac{2 \mu^2 H_0}{2\ell + 1} \\
 0 & 0 & 0 & 2\mu(\gamma + \omega) + V_b & -\dfrac{2 \mu^2 H_0}{2\ell + 1} \\
 4 \pi \mu^2 \ell H_0 & 4 \pi \mu^2 \ell H_0 & -4 \pi \mu^2 (\ell+1) H_0 & -4 \pi \mu^2 (\ell+1) H_0 & V - 2 \mu^2 U_0 
\end{pmatrix}$}.
\end{equation}

Here, the effective radial potentials are defined as
\begin{align}
    V_a(r) &\equiv \frac{\ell(\ell-1)}{r^2} + 2 \mu^2 U_0 \,, \\
    V_b(r) &\equiv  \frac{(\ell+1)(\ell+2)}{r^2} + 2 \mu^2 U_0 \,,\\
    V(r) &\equiv \frac{\ell(\ell+1)}{r^2} + 2 \mu^2 U_0 \,. 
\label{align:effect-potential-scalar}
\end{align}
This formulation, i.e., expressing the perturbed master system in terms of the functions $\{ q_a, q_b \}$ rather than the polar functions $\{ Z,Z_\mathcal{P} \}$, provides direct access to the intrinsic spin properties of the hedgehog NPS. Notice that $V_a$ and $V_b$ are the effective potentials for the corresponding perturbation functions $q_a$ and $q_b$. To interpret these in terms of an intrinsic spin, we can define a generic spin-dependent potential in terms of the total angular momentum $j=s+\ell$:
\begin{equation}
     V_s \equiv \frac{j (j+1)}{r^2} + 2 \mu^2 U_0 \,,
\end{equation}
where $s$ denotes the spin projection. Setting $s=-1$ and $s=+1$, we recover the corresponding effective potentials $V_a$ and $V_b$. Thus, the perturbation functions $q_a$ and $q_b$ encode the two transverse degrees of freedom associated with the spin projections $s=-1$ and $s=+1$, respectively. 

The solution to the inhomogeneous system~\eqref{eq:inhomogeneous-Matrix-scalar}
is obtained by variation of parameters (see, e.g., Ref.~\cite{boyce2004}),
\begin{equation}
    \mathbf{X} = \mathbf{F} \int \dd r\, \mathbf{F}^{-1} \mathbf{P} \,,
\label{eq:var-param-scalar-matrix}
\end{equation}
where $\mathbf{F}(r)$ is the fundamental matrix and $\mathbf{P}(r)$ is the source vector. The fundamental matrix is constructed from a complete basis of linearly independent
solutions of the homogeneous system
\begin{equation}
\partial_r \mathbf{X} - V_B(r) \mathbf{X} = \mathbf{0} \,,
\label{eq:fun-matrix-homog-system}
\end{equation}
satisfying the appropriate boundary conditions.
Following the approach used for BSs in Refs.~\cite{Macedo_2013,Annulli:2020l,Macedo:2026},
we impose regularity at the origin and the Sommerfeld radiation condition at spatial infinity. 

To implement the boundary conditions at the origin, we apply the Frobenius method to
the homogeneous perturbation system~\eqref{eq:master-hedge-poisson-qa-qb}--\eqref{eq:master-hedge-qb}.
Keeping only the leading term in each expansion, with real constant $u_0$ and complex
constants $a_{0k}$ and $b_{0k}$, we obtain\footnote{When numerically integrating the equations, we typically start from $r_0=10^{-3}M$.}

\begin{equation}
    \begin{split}
        q_{a_k}(r \to 0) &\sim a_{0k} \,r^\ell \,,\\
        q_{b_k}(r \to 0) &\sim b_{0k} \,r^{\ell+2} \,,
        \quad \quad k=1,2\\
        u(r \to 0) &\sim u_0 \,r^{\ell+1} \,.
    \end{split} \label{eq:boundary-qa-qb-u-r0}
\end{equation}
At spatial infinity, we impose the Sommerfeld radiation condition by considering
purely outgoing waves for the Proca perturbations. On the other hand, since the radial hedgehog function $H_0$ vanishes at infinity, the Poisson equation~\eqref{eq:master-hedge-poisson-qa-qb}
becomes a homogeneous ODE at large radii, and from
Eqs.~\eqref{eq:master-hedge-poisson-qa-qb}--\eqref{eq:master-hedge-qb},
we obtain\footnote{When numerically integrating the equations, we consider a sufficiently large radius that guarantees numerical convergence and that satisfies $r_\infty \gg R_\text{\tiny NPS}$.}
\begin{equation}
    \begin{split}
        q_{a_k}(r \to \infty) &\sim q_{a_k}^\infty \,e^{\mathrm{i} k_{\scalebox{0.4}{$k$}} r} r^{\mathrm{i} \frac{  M \mu^2}{k_{\scalebox{0.35}{$k$}}} } \,,\\
        q_{b_k}(r \to \infty) &\sim q_{b_k}^\infty \,e^{\mathrm{i} k_{\scalebox{0.4}{$k$}} r} r^{\mathrm{i} \frac{  M \mu^2}{k_{\scalebox{0.35}{$k$}}} } \,,\\
        u(r \to \infty) &\sim u_\infty \, r^{-\ell} \,,
    \end{split}\label{eq:boundary-qa-qb-u-rinf}
\end{equation}
for $k=1,2$, where $k_k$ are the corresponding wave numbers, $q_{a_k}^\infty$ and $q_{b_k}^\infty$ are complex asymptotic amplitudes, and
$u_\infty$ is real. For the Proca perturbations, the expressions above are kept up to
order $\mathcal{O}(r^0)$ in the asymptotic expansion. In the Newtonian regime, where $\gamma \ll \mu$, we have
\begin{equation}
    \begin{split}
        k_1 &\equiv \sqrt{2 \mu ( \omega - \gamma)} \,,\\
        k_2 &\equiv -\left( \sqrt{-2 \mu ( \omega + \gamma)} \right)^* \,,\\
    \end{split}\label{eq:k1-k2}
\end{equation}
using the principal complex square-root in the second expression. 

The complete set of boundary conditions at the origin and at infinity can be represented as
\begin{equation}
    \begin{split}
        &\mathbf{X}(r\to r_b) \sim \\
        &\lim_{r\to r_b} \Big( q_{a_1}, q_{a_2}, q_{b_1}, q_{b_2}, u, \partial_r q_{a_1}, \partial_r q_{a_2}, \partial_r q_{b_1}, \partial_r q_{b_2}, \partial_r u \Big)^T ,
    \end{split} \label{eq:X-vector-FunMat}
\end{equation}
where $r_b=\{0,+\infty\}$. To construct the fundamental matrix, we use the canonical basis
$\mathcal{B} = \{ \mathbf{e_1},\mathbf{e_2},\mathbf{...},\mathbf{e_9},\mathbf{e_{10}} \} $, where
$\mathbf{e}_{\boldsymbol{i}}$ is a ten-dimensional column vector with unit entry in the $i$th
position and zeros elsewhere. We introduce the set of linearly independent homogeneous
solutions
$\{\mathbf{Z}_\mathbf{(1)}, \mathbf{Z}_\mathbf{(2)},\textbf{...} ,  \mathbf{Z}_\mathbf{(9)}, \mathbf{Z}_\mathbf{(10)} \}$ that satisfy the boundary conditions
\begin{equation}
    \begin{split}
    \mathbf{Z}_{\boldsymbol{(i)}}(r \to 0) &\sim  \lim_{r \to 0} \Big( \mathrm{X}_i \,\mathbf{e}_{\boldsymbol{i}} +  \mathrm{X}_{i+5} \,\mathbf{e}_{\boldsymbol{i+5}} \Big) \,,\\
    \mathbf{Z}_{\boldsymbol{(i+5)}}(r \to \infty) &\sim  \lim_{r \to \infty} \Big( \mathrm{X}_i \,\mathbf{e}_{\boldsymbol{i}} +  \mathrm{X}_{i+5} \,\mathbf{e}_{\boldsymbol{i+5}} \Big) \,,
    \end{split}
\end{equation}
for $\boldsymbol{i}=1,2,3,4,5$, where $\mathrm{X}_i$ denotes the $i$th component of $\mathbf{X}(r \to r_b)$ in Eq.~\eqref{eq:X-vector-FunMat}. The fundamental matrix is then given by
\begin{equation}
    \mathbf{F}(r) \equiv \big( \mathbf{Z}_\mathbf{(1)}, \mathbf{Z}_\mathbf{(2)},\mathbf{Z}_\mathbf{(3)}, \mathbf{...}, \mathbf{Z}_\mathbf{(8)}, \mathbf{Z}_\mathbf{(9)}, \mathbf{Z}_\mathbf{(10)} \big)\,.
\label{eq:funMatrix-NPS-hedge}
\end{equation}
The solutions $\mathbf{Z}_{\boldsymbol{(i)}}$ are obtained by integrating the homogeneous system
\eqref{eq:fun-matrix-homog-system} from the origin, with the boundary
conditions~\eqref{eq:boundary-qa-qb-u-r0}, whereas the solutions
$\mathbf{Z}_{\boldsymbol{(i+5)}}$ are obtained by integrating from infinity with the boundary
conditions~\eqref{eq:boundary-qa-qb-u-rinf}.

Finally, from Eq.~\eqref{eq:var-param-scalar-matrix}, we obtain the solution
$\{q_{a_1},q_{a_2},q_{b_1},q_{b_2},u\}$. We note, however, that to compute the energy loss rate
in Eq.~\eqref{eq:Eloss-expression2}, only the perturbation functions $q_{a_1}(r)$ and $q_{b_1}(r)$ are needed. Those are given explicitly by:
\begin{equation}
\begin{split}
q_{x}(r) &= 4\pi \Bigg[
\sum_{n=1}^{5} F_{x_i,n}(r) \int_{\infty}^{r} \dd r' \, F_{n,10}^{-1}(r') p(r') \\
&\qquad+ \sum_{n=6}^{10} F_{x_i,n}(r) \int_{0}^{r} \dd r' \, F_{n,10}^{-1}(r') p(r')
\Bigg] \,,
\end{split} \label{eq:qa1-var-param}
\end{equation}
where $x\in\{a_1,b_1\}$, and $x_i$ denotes the corresponding component index:
$x_i=1$ for $x=a_1$, and $x_i=3$ for $x=b_1$. Henceforth, we use the compact notation
$q_x(r)$ for $x\in\{a_1,b_1\}$.

To obtain the Proca energy and Noether charge fluxes, only the asymptotic amplitudes at $r\to \infty$ are needed. Denoting those by
$q_{a_1}^\infty$ and $q_{b_1}^\infty$, from Eqs.~\eqref{eq:qa1-var-param} and~\eqref{eq:source-p}, we obtain 
\begin{equation}
\begin{aligned}
        q_{x}^\infty &= 4\pi \int_{0}^{\infty} \dd r' \, F_{x_i+5,10}^{-1}(r') p(r')  \\
        &= 4 \pi \tilde{p} \,F_{x_i+5,10}^{-1}(r_\text{orb},\omega) \delta(\omega-m\omega_\text{orb})\,, 
        \end{aligned} 
\end{equation}
with $\tilde{p} \equiv m_p\sqrt{\frac{\pi}{2}} \frac{Y^{\ell m} \left(\pi/2,0 \right)}{r_\text{orb}}$. 
As discussed in Appendix~\ref{app:Erad-Qrad-flux}, due to the integration over the frequency in
Eq.~\eqref{eq:fourier-pert-VSH}, the amplitudes only need to be evaluated at $\omega=m\omega_\text{orb}$, giving
\begin{equation}
    q_{x}^\infty = 4 \pi \tilde{p} \,F_{x_i+5,10}^{-1}(r_\text{orb},m\omega_\text{orb})\,.
\end{equation}
In what follows, we write the asymptotic amplitudes in this form.

To compute the energy and Noether charge fluxes, we also need to relate $q_{x}^\infty$ with the amplitudes $Z_1^{\infty}$ and $Z_{1\mathcal{P}}^{\infty}$. These can be obtained from Eq.~\eqref{eq:transformation-qa-qb} and read
\begin{equation}
\begin{aligned}
    &Z_1^\infty = \ell \,q_{a_1}^\infty - (\ell+1) \,q_{b_1}^\infty \,, \\
    &Z_{1\mathcal{P}}^\infty = q_{a_1}^\infty + q_{b_1}^\infty \,.
\end{aligned}
\end{equation}

Hereafter, we
refer to the procedure we just described as the full-setup method.

\subsubsection{Source-dominated approximation}
\label{subsubsec:Approx-hedge-EMRIs}

In the Newtonian limit, the NPS density is typically sufficiently small that the term $ \mu^2 \mathcal{A}_{0j} \delta \mathcal{A}^j $ in Eq.~\eqref{eq:SP-pert-poisson-source} is subleading\footnote{The same discussion of course applies to the term $ \mu^2 \mathcal{A}_{0j} \delta \mathcal{\bar{A}}^j $.} with respect to the source term $ P $ (see Refs.~\cite{Dyson:2025,Annulli:2020l} for a related discussion). Let $\varepsilon_\text{\tiny B}$ denote the field amplitude, such that
$\mathcal{A}_{0j}\sim\varepsilon_\text{\tiny B}$, and let $\varepsilon=m_p/\Mnps$ be the mass ratio. If the perturbation scales as $ \delta \mathcal{A}^j \sim \varepsilon_\text{\tiny B} \varepsilon $, then $ \mathcal{A}_{0j} \delta \mathcal{A}^j \sim \varepsilon_\text{\tiny B}^2 \varepsilon $, whereas the source term scales as $P\sim\varepsilon$. Thus, the perturbed Poisson equation~\eqref{eq:SP-pert-poisson-source} schematically reads
\begin{equation}
\begin{split}
\nabla^2 \delta U &\sim \mathcal{A}_{0j} \delta \mathcal{A}^j + P \\ 
&\sim \varepsilon_\text{\tiny B}^2 \varepsilon + \varepsilon \,. 
\end{split}
\end{equation}
Since $\varepsilon_\text{\tiny B}\ll1$ for NPSs, the field interaction term is suppressed by
$\varepsilon_\text{\tiny B}^2$ and is therefore subleading. The Poisson equation then reduces to
\begin{equation}
    \nabla^2\delta U \simeq 4 \pi P \,,
\label{eq:poisson-approx}
\end{equation}
or, equivalently,
\begin{equation}
  \hat{\mathcal{D}}_0 u(r) = 4 \pi p(r) \,,
\label{eq:poisson-approx-masterr}
\end{equation}
where $p(r)$ is the source term for a circular, equatorial orbit, defined in
Eq.~\eqref{eq:source-p}. Applying variation of parameters to
Eq.~\eqref{eq:poisson-approx-masterr}, we obtain the solution
\begin{align}
    u &(r) = - (2\pi)^{3/2} m_p \frac{Y^{\ell m}\left( \frac{\pi}{2}, 0 \right)}{2\ell + 1} \delta(\omega - m\omega_\text{orb}) \label{eq:u-general}\\
& \times \left[ \left( \frac{r}{r_\text{orb}} \right)^{-\ell} \Theta(r - r_\text{orb}) 
+ \left( \frac{r}{r_\text{orb}} \right)^{\ell+1} \Theta(r_\text{orb} - r) \right]\,, \notag
\end{align}
where $\Theta(x)$ is the Heaviside step function. 

From the polar sector hedgehog master equations
\eqref{eq:master-hedge-qa}--\eqref{eq:master-hedge-qb}, evaluated at
$\omega=m\omega_\text{orb}$ for the reasons discussed above, we obtain the completely decoupled second-order ODE system:
\begin{equation}
    \begin{split}
    &\partial_r^2 q_{a_1} + \left( 2\mu m \omega_{\text{orb}} - \frac{\ell(\ell-1)}{r^2} + \xi(r) \right) q_{a_1}  = S_{a_1}(r) \,,\\
    &\partial_r^2 q_{b_1} + \left( 2\mu m \omega_{\text{orb}} - \frac{(\ell+1)(\ell+2)}{r^2} + \xi(r) \right) q_{b_1}  = S_{b_1}(r) \,,  
    \end{split} 
    \label{eq:NDSolve-eqs-master}
\end{equation}
with $\xi(r)=- 2\mu(\gamma + \mu U_0(r))$ and the inhomogeneous source terms given by $S_{a_1}(r)=-S_{b_1}(r) = \frac{2 \mu^2 H_0 u(r)}{2\ell+1}$. 

The polar equations for $q_x(r)$, with $x\in\{a_1,b_1\}$, can again be solved via variation of parameters but now, since $u(r)$ is given by Eq.~\eqref{eq:u-general}, the solutions simplify considerably. Let $\{q_x^0(r),q_x^\infty(r)\}$ denote two linearly independent homogeneous solutions satisfying the boundary conditions at the origin and at infinity. The general solutions are then
\begin{equation}
\begin{aligned}
    q_{x}(r) &= \,q_{x}^\infty (r) \int_0^r \dd r' \,\frac{q_{x}^0 (r') S_{x}(r')}{W} \\
        & \qquad \qquad + q_{x}^0 (r) \int_r^\infty \dd r' \, \frac{q_{x}^\infty (r') S_{x}(r')}{W} \,,
\label{eq:var-parm-qa-qb-NDSolve}
\end{aligned}
\end{equation}
where 
\begin{equation}
    W \left[q_{x}^0(r) ,q_{x}^\infty(r)  \right] = q_{x}^0 (r) \frac{\dd q_{x}^\infty (r)} {\dd r} - q_{x}^\infty (r) \frac{\dd q_{x}^0 (r)} {\dd r} 
\end{equation}
denotes the Wronskian, constant\footnote{By Abel’s theorem \cite{boyce2004}, the Wronskian of a second-order ODE remains constant throughout its domain in the absence of a first-derivative term.} in $r$. The corresponding asymptotic amplitudes are then given by
\begin{equation}
    q_{x}^\infty \equiv q_{x}^\infty(r \to \infty) \int_0^\infty \dd r' \frac{q_{x}^0(r') \,S_{x}(r')}{W} \,,
\label{eq:NDSolve-var-qa-qbbb}
\end{equation}
where $q_{x}^\infty \equiv q_x(r \to \infty)$. We compute the homogeneous solutions $q_x^0(r)$ and $q_x^\infty(r)$ numerically, by
direct integration of the sourceless version of Eq.~\eqref{eq:NDSolve-eqs-master},
i.e., by setting $S_{a_1}(r)=S_{b_1}(r)=0$. In each case, the integration is performed with
the appropriate boundary conditions: regularity at the origin and Sommerfeld radiation condition at infinity, as detailed in the previous section.
\\

\noindent \textbf{High-frequency approximation.} In the high-frequency limit $\gamma, \mu U_0 \ll \omega_\text{orb}$, the system
\eqref{eq:NDSolve-eqs-master} reduces to
\begin{equation}
\begin{split}
    &\partial_r^2 q_{a_1} + \left( 2\mu m \omega_{\text{orb}} - \frac{\ell(\ell-1)}{r^2} \right) q_{a_1}  = S_{a_1}(r) \,,\\
    &\partial_r^2 q_{b_1} + \left( 2\mu m \omega_{\text{orb}} - \frac{(\ell+1)(\ell+2)}{r^2} \right) q_{b_1}  = S_{b_1}(r)\,.
\end{split}
\end{equation}
In this limit, the corresponding homogeneous solutions can be written analytically as
\begin{equation}
    \begin{aligned}
        q_{x}^0(r) &= \sqrt{r} \, J_{\ell + 1/2 + s_{x}} \left( \sqrt{2 \mu m \omega_{\text{orb}}} \, r \right) \,,\\
        q_{x}^\infty(r) &= \sqrt{r} \, H^{(1)}_{\ell + 1/2 + s_{x}} \left( \sqrt{2 \mu m \omega_{\text{orb}}} \, r \right)\,,
    \end{aligned}
\label{eq:Z-homogs-approx}
\end{equation}
where
\begin{equation}
s_{x} = 
\begin{cases}
-1 & \quad \text{if } \,\,q_x = q_{a_1}\,, \\
+1  & \quad \text{if } \,\,q_x = q_{b_1}\,.
\end{cases}
\end{equation}
Here, $J_\nu (x)$ and $H_\nu^{(1)} (x)$ are the Bessel and Hankel functions of the first kind~\cite{abramowitz-Bessel}. The asymptotic behaviors of the outgoing solutions $q_{x}^\infty(r)$ are then given by
\begin{equation}
    q_{x}^\infty(r\to\infty) \simeq (-\mathrm{i})^{\ell+1+s_{x}} \sqrt{\frac{2}{\pi}} \frac{e^{\mathrm{i}\sqrt{2\mu m \omega_\text{orb}} r} }{(2 \mu m \omega_\text{orb})^{1/4}} \,.
\label{eq:Z-homogs-approx-asympt}
\end{equation}
In this regime, the Wronskian remains constant and is given by $W\left[q_{x}^0(r),q_{x}^\infty(r)\right] = \frac{2 \mathrm{i}}{\pi}$. The amplitudes $q_{x}^\infty$ are then obtained from
Eq.~\eqref{eq:NDSolve-var-qa-qbbb}, using the analytical homogeneous solutions
\eqref{eq:Z-homogs-approx}.

To check the accuracy of our numerical results, below we will compare the results obtained using the full-setup method and the approximations we just described.

\subsection{Ground-state configuration}
\label{subsec:NPS-gs}
Unlike for the hedgehog configuration, the master system of equations~\eqref{eq:master-poisson}--\eqref{eq:master-proca3}, describing perturbations of the NPS ground-state configuration, effectively contains an infinite number of equations due to the coupling between different $\ell$ modes, as described in
Sec.~\ref{subsec:gs-NPS-small}. Another important difference from the hedgehog case is that now the axial perturbations are coupled to the polar perturbations through Eqs.~\eqref{eq:master-proca2} and \eqref{eq:master-proca3}. Thus, the axial sector also needs to be included. 

To solve this system for a given azimuthal number $m$, we truncate the axial sector at $\ell=L$ and the polar sector at $\ell=L+1$ by imposing
\begin{equation}
    Z_{k\mathcal{A}}^\ell=0\,, \quad Z_k^{\ell+1}=Z_{k\mathcal{P}}^{\ell+1}=0\,, \quad \text{for} \quad \ell \geq L\,,
\end{equation}
with $k=1,2$. Moreover, since $\ell\geq |m|$, we can set all modes with
$\ell<|m|$ to zero. The resulting truncated system contains at most $7L-2$ ODEs.
We can now apply the same framework as in the previous section to this truncated system of equations.

Including the point-particle's source term in Eq.~\eqref{eq:master-poisson}, the ground-state master system is written in the matrix
form of Eq.~\eqref{eq:inhomogeneous-Matrix-scalar}, with perturbation vector 
\begin{equation}
\begin{aligned}
    &\mathbf{X}= \\
    &\big( Z^\ell_1,Z^\ell_2,Z_{1\mathcal{P}}^\ell,Z_{2\mathcal{P}}^\ell,Z_{1\mathcal{A}}^\ell,Z_{2\mathcal{A}}^\ell, u^\ell ,\text{...}, Z^L_1,Z^L_2,Z_{1\mathcal{P}}^L,Z_{2\mathcal{P}}^L,u^L, \\
    &\partial_r Z^\ell_1,\partial_r Z^\ell_2, \partial_r Z_{1\mathcal{P}}^\ell,\text{...},\partial_r Z^L_1,\partial_r Z^L_2,\partial_r Z_{1\mathcal{P}}^L,\partial_r Z_{2\mathcal{P}}^L,\partial_r u^L \big)^T    ,
\end{aligned}
\end{equation}
source vector
\begin{equation}
    \mathbf{P}= \big(0,0,0,0,0,0,4\pi p^{\ell},\text{...},0,4 \pi p^{L-1}, 0,0,0,0,4\pi p^{L} \big)^T ,
\end{equation}
and matrix $V_B$, whose dimension depends on the value of $m$ and the chosen truncation $L \geq m$. The number of variables $\{Z^\ell_1,Z^\ell_2,Z_{1\mathcal{P}}^\ell,Z_{2\mathcal{P}}^\ell,Z_{1\mathcal{A}}^\ell,Z_{2\mathcal{A}}^\ell, u^\ell\}$, for $m \leq \ell \leq L$, is given by $N=7(L-m)+5$. Including the first derivatives, the dimensionality is given by $\mathrm{dim}\equiv2N = 14(L-m)+10$. 

To simplify the system and improve computational efficiency, we apply the linear
transformation~\eqref{eq:transformation-qa-qb} to the polar sector and work with the functions
$q_{a_k}^\ell$ and $q_{b_k}^\ell$. The perturbation vector becomes
\begin{equation}
\begin{aligned}
    &\mathbf{X}= \\
    &\big( q_{a_1}^\ell,q^\ell_{a_2},q_{b_1}^\ell,q_{b_2}^\ell,Z_{1\mathcal{A}}^\ell,Z_{2\mathcal{A}}^\ell, u^\ell ,\text{...},q_{a_1}^L,q_{a_2}^L,q_{b_1}^L,q_{b_2}^L,u^L, \\
    &\partial_r q_{a_1}^\ell,\partial_r q_{a_2}^\ell,\text{...},\partial_r q_{a_1}^L,\partial_r q_{a_2}^L,\partial_r q_{b_1}^L,\partial_r q_{b_2}^L,\partial_r u^L \big)^T   \,.
\end{aligned}
\end{equation}

It is useful to define the positions of the perturbations $q_{a_1}^\ell$, $q_{b_1}^\ell$ and $Z_{1\mathcal{A}}^\ell$ in the vector $\mathbf{X}$, which are given by
\begin{equation}
\begin{aligned}
    \mathrm{pos}(q_{a_1}^\ell) &= 7(\ell-m)+1\,,
    \quad && m\leq \ell \leq L\,,\\
    \mathrm{pos}(q_{b_1}^\ell) &= 7(\ell-m)+3\,,
    \quad && m\leq \ell \leq L\,,\\
    \mathrm{pos}(Z_{1\mathcal{A}}^\ell) &= 7(\ell-m)+5\,,
    \quad && m\leq \ell \leq L-1 \,.
\end{aligned}
\end{equation}
Similarly, the nonzero entries of $\mathbf{P}$ are at position
\begin{equation}
    \mathrm{pos}(p^\ell) = 7(\ell - m + 1) - 2 \delta_{\ell, L}   \,,
\end{equation}
where $m\leq \ell \leq L$, and $\delta_{\ell,L}$ is the Kronecker delta. 

Defining the fundamental matrix, as described for the hedgehog case, and applying the method of variation of parameters then yields
\begin{equation}
\begin{aligned}
q_{x}^{\ell^*}(r) &= 4\pi \sum_{\ell = m}^{L} \sum_{n=1}^{2N}
F_{\mathrm{pos}(q_{x}^{\ell^*}), n}(r)
\int_{r_b}^{r} \dd r' F^{-1}_{n,2N}(r')p^\ell(r')\,, \\
Z_{1\mathcal{A}}^{\ell^*}(r) &= 4\pi \sum_{\ell = m}^{L-1} \sum_{n=1}^{2N}
F_{\mathrm{pos}(Z_{1\mathcal{A}}^{\ell^*}), n}(r)
\int_{r_b}^{r} \dd r' F^{-1}_{n,2N}(r') p^\ell(r')\,, 
\end{aligned}
\label{eq:gs-var-param-solutions}
\end{equation}
with
\begin{equation}
r_b =
\begin{cases}
\infty\,, & n \leq N\,, \\
0\,, & n > N \,,
\end{cases}
\end{equation}
where, as in the previous section, $q_x^\ell(r)$ denotes either $q_{a_1}^\ell(r)$ or $q_{b_1}^\ell(r)$. Here, $\ell^*$ denotes the multipole of
interest, with $m \leq \ell^* \leq L$ for the polar sector and
$m \leq \ell^* \leq L-1$ for the axial sector. The corresponding asymptotic amplitudes are 
\begin{equation}
\begin{aligned}
    q_{x}^{\infty,\ell^*} 
    &= 4\pi \sum_{\ell=m}^{L} \tilde{p}^{\,\ell} F_{N+\mathrm{pos}(q_{x}^{\ell^*}),N+\mathrm{pos}(p^\ell)}^{-1}(r_\text{orb},m \omega_\text{orb}) \,,\\
    Z_{1\mathcal{A}}^{\infty,\ell^*} 
    &= 4\pi \sum_{\ell=m}^{L-1} \tilde{p}^{\,\ell} F_{N+\mathrm{pos}(Z_{1\mathcal{A}}^{\ell^*}),N+\mathrm{pos}(p^\ell)}^{-1}(r_\text{orb},m \omega_\text{orb}) \,,
\end{aligned}
\end{equation}
with $\tilde{p}^{\ell}=m_p \sqrt{\frac{\pi}{2}} \,\frac{Y^{\ell m}(\pi/2,0)}{r_\text{orb}}$. Finally, to compute the energy fluxes, we again use the linear transformation~\eqref{eq:transformation-qa-qb} to obtain the polar amplitudes.

\subsubsection{Source-dominated approximation}
\label{subsubsec:Approx-GS-EMRIs}

As discussed in the hedgehog case, it is useful to take the approximation in which the solution to the potential $u^\ell(r)$ is given, Eq.~\eqref{eq:u-general}, such that $u^\ell(r)$ only appears as a source term in Eqs.~\eqref{eq:master-proca1}--\eqref{eq:master-proca3}. Following the same steps as in the hedgehog case, we find that, for any truncation $L$, we can completely decouple the system of equations and obtain
\begin{equation}
    \begin{split}
    &\partial_r^2 q_{a_1}^\ell + \left( 2\mu m \omega_{\text{orb}} - \frac{\ell(\ell-1)}{r^2} + \xi(r) \right) q_{a_1}^\ell  = \tilde S^\ell_{a_1}(r) \,,\\
    &\partial_r^2 q_{b_1}^\ell + \left( 2\mu m \omega_{\text{orb}} - \frac{(\ell+1)(\ell+2)}{r^2} + \xi(r) \right) q_{b_1}^\ell  = \tilde S^\ell_{b_1}(r) \,,\\
    &\partial_r^2 Z_{1\mathcal{A}}^\ell + \left( 2\mu m \omega_{\text{orb}} - \frac{\ell(\ell+1)}{r^2} + \xi(r) \right) Z_{1\mathcal{A}}^\ell  =  \tilde S^\ell_{\mathcal{A}}(r)  \,. 
    \end{split} \label{eq:NDSolve-eqs-gs}
\end{equation}
Note that, in contrast with the hedgehog case, the equation for the axial sector is now also inhomogeneous. The source terms $\tilde S^\ell_{a_1}(r)$, $\tilde S^\ell_{b_1}(r)$, and $\tilde S^\ell_{\mathcal{A}}(r)$ involve contributions proportional to $2\mu^2 h_0$, with sums over $\mathcal{Q}_\ell$ and $u^\ell$. More explicitly: (i) the equation for $q_x^{\ell \,\text{even}}$ contains a sum over $u^{\ell\,\text{odd}}$; (ii) the equation for $Z_{1\mathcal{A}}^{\ell\,\text{even}}$ contains a sum over $u^{\ell\,\text{even}}$; (iii) and vice versa. 

Each equation in the system~\eqref{eq:NDSolve-eqs-gs} can be solved through variation of parameters, allowing us to obtain the asymptotic amplitudes for a chosen multipole $\ell^*$
\begin{equation}
\begin{aligned}
    q_{x}^{\infty,\ell^*} &= q_{x}^{\infty,\ell^*}(r\to \infty) \int_0^\infty \dd r' \frac{q_{x}^{0,\ell^*}(r') \,\tilde S^\ell_{x}(r')}{W}\,,\\
    Z_{1\mathcal{A}}^{\infty,\ell^*} &= Z_{1\mathcal{A}}^{\infty,\ell^*}(r\to \infty) \int_0^\infty \dd r' \frac{Z_{1\mathcal{A}}^{0,\ell^*}(r') \,\tilde S^\ell_{\mathcal{A}}(r')}{W}\,,
\end{aligned}\label{eq:var-paramsssssss}
\end{equation}
where $x\in\{a_1,b_1\}$, and the homogeneous solutions
$q_x^{0,\ell^*}(r)$ and $q_x^{\infty,\ell^*}(r)$, as well as $Z_{1\mathcal{A}}^{0,\ell^*}(r)$ and
$Z_{1\mathcal{A}}^{\infty,\ell^*}(r)$, are computed by direct integration of the sourceless version of Eq.~\eqref{eq:NDSolve-eqs-gs} with the appropriate boundary conditions at the origin and at infinity, respectively. 
\\

\noindent\textbf{High-frequency approximation.} In the limit $\gamma, \mu U_0 \ll \omega_\text{orb}$, Eq.~\eqref{eq:NDSolve-eqs-gs} reduces to
\begin{equation}
    \begin{split}
    &\partial_r^2 q_{a_1}^\ell + \left( 2\mu m \omega_{\text{orb}} - \frac{\ell(\ell-1)}{r^2}  \right) q_{a_1}^\ell  = \tilde S^{\ell}_{a_1}(r) \,,\\
    &\partial_r^2 q_{b_1}^\ell + \left( 2\mu m \omega_{\text{orb}} - \frac{(\ell+1)(\ell+2)}{r^2}  \right) q_{b_1}^\ell  = \tilde S^{\ell}_{b_1}(r) \,,\\
    &\partial_r^2 Z_{1\mathcal{A}}^\ell + \left( 2\mu m \omega_{\text{orb}} - \frac{\ell(\ell+1)}{r^2} \right) Z_{1\mathcal{A}}^\ell  =  \tilde S^{\ell}_{\mathcal{A}}(r)   \,.
    \end{split}
\end{equation}
In this limit, we can then obtain the asymptotic amplitudes~\eqref{eq:var-paramsssssss} using the analytical homogeneous solutions given in~\eqref{eq:Z-homogs-approx}, where $Z_{1\mathcal{A}}^{0/\infty}(r)$ also takes the same form but with $s_{x}=0$.


\section{Results}
\label{sec:numerical_results}

Using the methods developed in the previous section, we now study the orbital energy loss rates $\dot{E}^{\rm lost}$ for a point particle in a circular, equatorial orbit around an NPS. We consider two different scenarios: one in which a central BH of mass $M_{\text{\tiny BH}} \ll M$ is present, referred to as a ``parasitic BH'' following Ref.~\cite{Cardoso:2022nzc}, and another in which this central BH is absent. Here, $M$ denotes the mass of the NPS. The parasitic BH is modeled as a point source located at the center of the NPS. Because $M_{\text{\tiny BH}} \ll M$, we neglect its effect on the NPS configuration, namely absorption at the BH horizon~\cite{Annulli:2020l,Cardoso:2022nzc}. Including a central parasitic BH is motivated by its ability to mimic astrophysical systems such as a compact object inspiralling around the massive BH at a galactic center, with the NPS describing the core of the DM halo~\cite{Annulli:2020l}. Our main goal is to study how the rate of energy lost by the point particle differs between the two NPS configurations as the particle moves through them. Ultimately, we aim to compare these results with previous ones on scalar NBSs~\cite{Annulli:2020l,Duque_2024}, in order to understand how the response of the bosonic configurations depends on the intrinsic spin of the field.
\\

\noindent \textbf{(i) Methods and domains of applicability.} In order to check the validity of our results, we consider the three methods discussed in the previous section: the full-setup method (using the fundamental matrix), the source-dominated approximation, and the high-frequency approximation. 

The full-setup method involves no approximations and can be applied both with and without a central parasitic BH. In the absence of a BH, the gravitational potential is regular at the origin, and the method remains numerically stable down to small orbital radii. In the presence of a parasitic BH, the Newtonian potential becomes
\begin{equation}
    U_n(r) = U_0(r) - \frac{M_{\text{\tiny BH}}}{r}\,,
\end{equation}
where $U_0$ is the background potential of the NPS configuration and the term $- M_{\text{\tiny BH}}/r$ models the BH as a Newtonian point mass. The $1/r$ behavior of the BH term implies that $U_n(r)$ diverges for small enough radii. In this case, it becomes increasingly difficult to numerically invert the fundamental matrix at small radii. For this reason, we show the full-setup results only for orbital radii above a minimum threshold $r_{\rm thr}$, i.e., $r_{\rm orb}\gtrsim r_{\rm thr}$, where the inversion of the fundamental matrix is stable.

The source-dominated approximation can likewise be implemented with and without a parasitic BH. As we discuss below, it shows good agreement with the full-setup results since the Proca field is sufficiently dilute (see discussion in Sec.~\ref{subsubsec:Approx-hedge-EMRIs}). Unlike the full-setup method, this method remains numerically stable at very small radii even in the presence of a parasitic BH, since it does not require the inversion of the fundamental matrix. We therefore present its results over an orbital-radius domain that extends down to arbitrarily small radii $r_{\rm orb} \gtrsim 0$.

The high-frequency approximation is only applicable in the presence of a parasitic BH and at small enough orbital radii. From the condition for circular orbits, the orbital angular velocity at $r=r_{\rm orb}$ is given by
\begin{equation}
    \omega_{\text{orb}} = \sqrt{\frac{1}{r} \frac{\dd U_n}{\dd r}} \bigg|_{r = r_\text{orb}} \,.
    \label{eq:worb}
\end{equation}
On the other hand, in the presence of a parasitic BH, the Newtonian potential $U_n(r)$ diverges as $r\to 0$. This allows for orbital frequencies satisfying $\omega_{\text{orb}} \gg \gamma,\, \mu U_n(r)$, a condition that is met only at sufficiently small radii. Like the source-dominated approximation, the high-frequency approximation remains numerically stable at very small radii.

Overall, the three methods are complementary: the full-setup method is the most general, involving no approximations, but becomes numerically challenging at small radii; by contrast, the source-dominated and high-frequency approximations rely on specific physical approximations, yet remain numerically stable within their respective domains of validity. Comparisons between the different methods can be found in Appendix~\ref{subsec:3_methods}. Since those show an excellent agreement between all three methods, the results we discuss below use either the full-setup or the source-dominated approximation.
\\

\noindent \textbf{(ii) Truncation dependence.} As discussed in the previous section, to solve the system of perturbative equations for the ground-state NPS, we need to choose a truncation order $L$. Although applying the full-setup method is computationally demanding due to its high dimensionality, it can in principle be implemented for any $L$. For this method, we therefore choose a mode-dependent truncation $L$ that ensures numerical convergence while remaining computationally efficient (see Appendix~\ref{sec:convergence} for a detailed discussion on the convergence).

A similar truncation also needs to be done to compute the source terms of the source-dominated and high-frequency equations, as discussed in Sec.~\ref{subsubsec:Approx-GS-EMRIs}. For these methods, we use $L=10$ for all results shown below, which, as we discuss in Appendix~\ref{sec:convergence}, ensures numerical convergence.

\subsection{Energy losses in the presence of a central parasitic black hole and without it}
\label{subsec:w_wo_BH_numerical-results}

We start by comparing the energy loss rate\footnote{Here, $\dot{E}^{\rm lost}_{\ell m}$ denotes the contribution from a single $(\ell,m)$ mode, whose expression is obtained from Eq.~(\ref{eq:Eloss-expression2}) without the summation over $\ell$ and $m$.} $\dot{E}^{\text{lost}}_{\ell m}$ for inspirals into the ground-state and hedgehog NPS, in the presence of a central parasitic BH and without it. For comparison with previous results obtained with NBSs, we consider the same fiducial values for $M\mu$ as Ref.~\cite{Duque_2024} and fix the mass of the parasitic BH to $M_{\text{\tiny BH}}=0.02M$ (cf. Fig.~1 in~\cite{Duque_2024}). 

\begin{figure}[htb]
\includegraphics[width=0.45\textwidth]{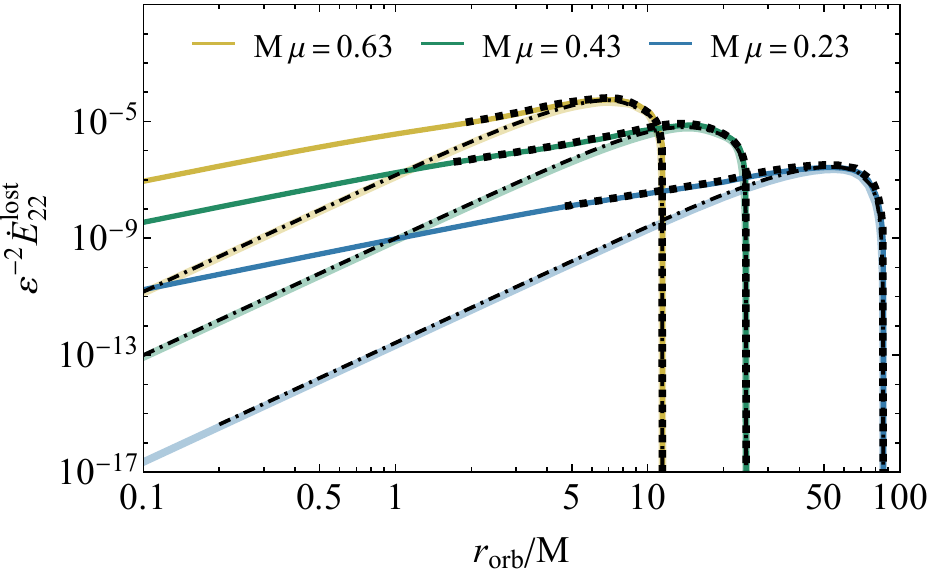}
\caption{Energy loss rate $\varepsilon^{-2}\dot{E}^{\rm lost}_{22}$ for the $\ell=m=2$ mode
in the ground-state NPS background as a function of the particle's orbital radius $r_{\rm orb}/M$, for three values of the coupling $M\mu$. Results obtained with the source-dominated approximation are shown as colored
solid lines (dark: with parasitic BH of mass $M_{\text{\tiny BH}}=0.02M$; light: without); results obtained with the full setup are shown as black lines (thick dashed: with parasitic BH; thin dot-dashed: without).}
\label{Fig:gs-Mus-l2m2-plot}
\end{figure}

\begin{figure}[htb]
\includegraphics[width=0.45\textwidth]{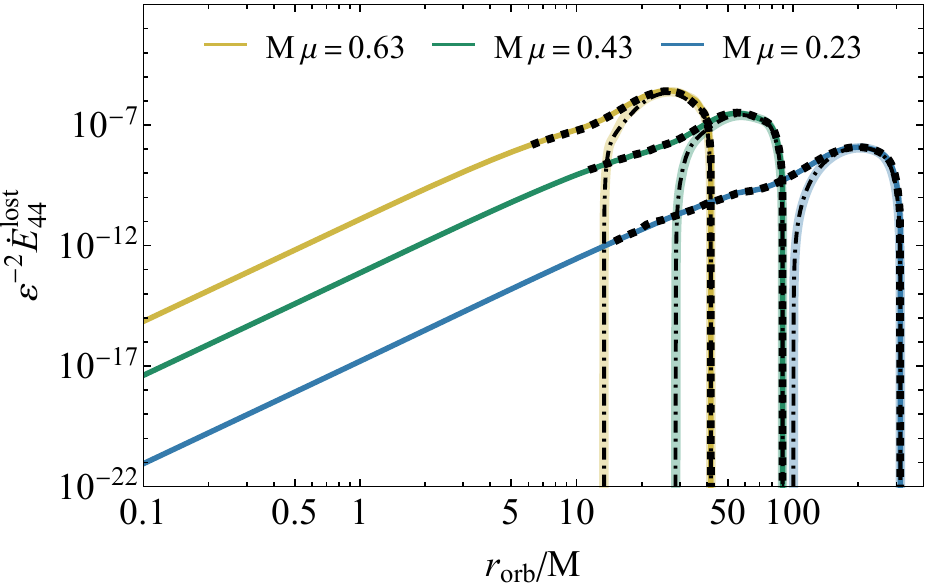}
\caption{Same as Fig.~\ref{Fig:gs-Mus-l2m2-plot}, but for a particle in circular orbit around the hedgehog NPS configuration, and here we instead show the energy loss rate $\varepsilon^{-2}\dot{E}^{\rm lost}_{44}$ for the $\ell=m=4$ mode.}
\label{Fig:hedge-Mus-l4m4-plot}
\end{figure}

For the ground-state configuration, we focus on the $\ell=m=2$ mode (Fig.~\ref{Fig:gs-Mus-l2m2-plot}), whereas for the hedgehog configuration, we consider $\ell=m=4$ (Fig.~\ref{Fig:hedge-Mus-l4m4-plot}); the reason for this mode choice is justified below. 

In all cases, we compare results obtained with the full setup and with the source-dominated approximation discussed in the previous section. For the full setup we only show results where the inversion of the fundamental matrix is stable. Overall, we find very good agreement between the two methods. This is expected given the dilute background profile, justifying the approximation employed in Sec.~\ref{subsubsec:Approx-hedge-EMRIs}. In Appendix~\ref{subsec:3_methods}, we also compare with results obtained with the high-frequency approximation discussed in the previous section. In this case, we also find excellent agreement in the regime where the approximation is valid, providing a robust benchmark for our numerical results.

For both ground-state and hedgehog NPSs, increasing the coupling $M\mu$ leads to more compact configurations. For $M\mu = 0.23$, $0.43$, and $0.63$, the radii enclosing $98\%$ of the total mass for the ground-state configuration are $R_{98} \simeq 172M$, $50M$, and $23M$, respectively. The corresponding values for the hedgehog configuration are $R_{98} \simeq 431M$, $124M$, and $58M$. The resulting increase in compactness (smaller $R_{98}$) enhances the interaction between the perturber and the surrounding medium, leading to higher maximum orbital energy losses as $M\mu$ increases. Comparing Figs.~\ref{Fig:gs-Mus-l2m2-plot} and~\ref{Fig:hedge-Mus-l4m4-plot}, we also see that the energy losses in the hedgehog configuration are overall smaller than the ground-state ones, with maximum values lower by roughly one order of magnitude. For a direct comparison using the same mode $\ell = m = 2$, the presence of a parasitic BH is required in the hedgehog case. This comparison is done in Appendix~\ref{subsec:3_methods}, where we show that, even in this case, the hedgehog configuration yields smaller overall energy losses, with maximum values one to two orders of magnitude below those of the ground-state configuration. This is expected given that, at fixed $M\mu$, hedgehog NPSs are less compact than the ground-state ones.

A qualitative difference between the two NPS classes appears in the hedgehog configuration when comparing the energy losses with and without a central parasitic BH. For a given $m>0$ mode, in the presence of a parasitic BH, the energy loss rate occurs over the interval $r_{\text{orb}} \in \left(\,0, r_{\text{cutoff}}\,\right)$, where $r_{\text{cutoff}}$ is set by the maximum orbital radius such that the argument of the square-root term in Eq.~\eqref{eq:Eloss-expression2} is positive. By contrast, in the absence of a parasitic BH, emission for a given $m$ mode is restricted to $r_{\text{orb}} \in \left(\,r_{\text{min}}, r_{\text{cutoff}}\,\right)$, as can be seen in Fig.~\ref{Fig:hedge-Mus-l4m4-plot}. The presence of a minimum orbital radius under which emission of Proca radiation is forbidden, in clear contrast with what happens for the ground-state NPS (cf. Fig.~\ref{Fig:gs-Mus-l2m2-plot}), is tied to the hedgehog field profile $H(r)$, which has a central node and a more diffuse distribution than the ground state (see Fig.~\ref{fig:bkg-hedgehog}). For $r_{\text{orb}} < r_{\text{min}}$ the gravitational potential is nearly flat, leading to $\omega_{\text{orb}}\simeq 0$ (cf. Eq.~\ref{eq:worb}). Consequently, the argument of the square-root factor appearing in
$\dot{E}^{\text{lost}}_{\ell m}$, i.e., $\sqrt{2\mu(m\omega_{\text{orb}} - \gamma)}$, becomes negative, suppressing radiation losses. For the hedgehog configuration, we verified that the condition $m\omega_{\text{orb}}>\gamma$, needed to obtain $\dot{E}^{\text{lost}}_{\ell m} \neq 0$, requires $m\geq 4$ for the parameters we consider, which motivates our focus on the lowest allowed mode, $\ell=m=4$. This is in contrast with the NPS ground state, for which the lowest allowed mode in the absence of a central parasitic BH is the $\ell=m=2$, for the parameters we consider, which is consistent with what happens for scalar NBSs~\cite{Macedo:2026}.

We should note that, although for $\ell=m\gg 1$ one finds $r_{\rm min}\to 0$, the corresponding fluxes are extremely small at low values of $r_{\rm orb}$. Thus, while higher modes can in principle allow emission at small radii even without a BH, their contribution is negligible for the cases considered here, and we do not include them in the present analysis. This behavior is in contrast to what happens in the ground-state configuration when no central parasitic BH is present. In this case, the field profile $h(r)$ is centrally peaked, yielding $\omega_{\text{orb}}>0$ throughout $r_{\text{orb}} \in (\,0,r_{\text{cutoff}}\,)$, and allowing emission for any $m \ge 2$.

In both the hedgehog and ground-state cases we see that the inclusion of a parasitic BH significantly enhances the energy loss at small $r_{\rm orb}$, while leaving the maximum values nearly unchanged. This enhancement is due to the additional $1/r$ contribution from the central BH, $U_n(r)=U_0(r)-\Mbh/r$, which deepens the gravitational potential well and therefore increases the orbital frequency $\omega_{\text{orb}}$ at small radii.
For all configurations considered here, the cutoff radius $r_{\text{cutoff}}$, above which the loss rates drop to zero in Figs.~\ref{Fig:gs-Mus-l2m2-plot} and \ref{Fig:hedge-Mus-l4m4-plot}, satisfies $r_{\text{cutoff}}<R_{98}$ and therefore lies well inside the NPS. For $r_{\rm orb}>R_{98}$, the background potential approaches $U_0(r)\sim -M/r$, with an additional contribution $-M_{\text{\tiny BH}}/r$ in the presence of a parasitic BH. The orbital frequency then becomes increasingly smaller as $r_{\rm orb}$ increases and only higher-$m$ modes can satisfy the condition $m\omega_{\text{orb}}-\gamma>0$, further suppressing emission at large orbital radii.

We close this part of the discussion by noting a curious feature of the loss rates which is realized for ground-state NPSs with and without a parasitic BH, and hedgehog NPSs with a parasitic BH. Although the energy density is largest near the center, the orbital energy loss rate does not peak there, and instead decreases sharply at small orbital radii. The same behavior also occurs in scalar BS configurations~\cite{Duque_2024,Macedo:2026}. Since $\omega_{\text{orb}}$ increases with smaller radii, one could have expected $\dot{E}^{\text{lost}}_{\ell m}$ to be the largest near the origin. A natural (conjectured) interpretation is that the gravitational potential well is also deepest in this region. As a result, the Proca energy radiated by the perturber may be more likely to remain trapped, reducing the amount of radiation that reaches infinity. At larger radii, the shallower potential well allows radiation to escape more efficiently, enhancing the loss rates and leading to the maximum values of $\dot{E}^{\text{lost}}_{\ell m}$ observed away from the center.

\subsection{Mode-by-mode contribution to the energy loss rate in the presence of a central parasitic black hole}
\label{subsec:with_a_parasitic_BH}

We now restrict to configurations with a central parasitic BH of mass $M_{\text{\tiny BH}}=0.02M$ and fix the field mass such that $M\mu=0.43$, for concreteness. Our final goal will be to compare the total orbital energy loss rate for a point particle in circular motion around the NPS configurations studied so far against the case of a scalar NBS.

\begin{figure}[htb]
\includegraphics[width=0.45\textwidth]{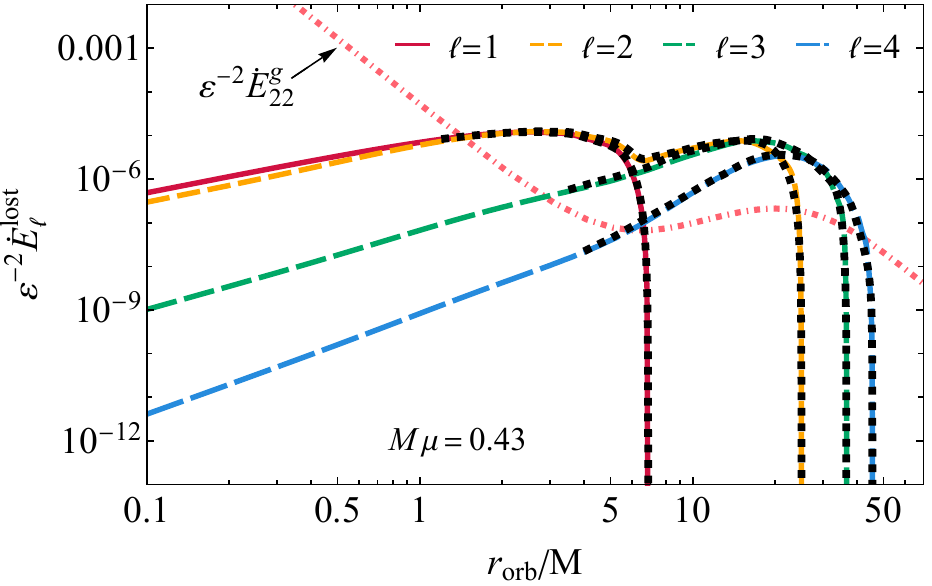}
\caption{Ground-state NPS. Energy loss rate $\varepsilon^{-2}\dot{E}^{\rm lost}_{\ell}$ as a function of the orbital radius $r_{\rm orb}/M$, for $M\mu=0.43$, in the presence of a central parasitic BH of mass $M_{\text{\tiny BH}}=0.02M$. We show the contribution of the four lowest multipoles $\ell=1,2,3,4$ to the Proca energy loss, obtained using the source-dominated approximation (colored solid/dashed lines) and using the full-setup method (thick black dashed lines). For comparison, the GW flux obtained using the quadrupole formula is also shown (light red dot-dashed line).}
\label{Fig:gs-Ls-mu43-plot}
\end{figure}

\begin{figure}[htb]
\includegraphics[width=0.45\textwidth]{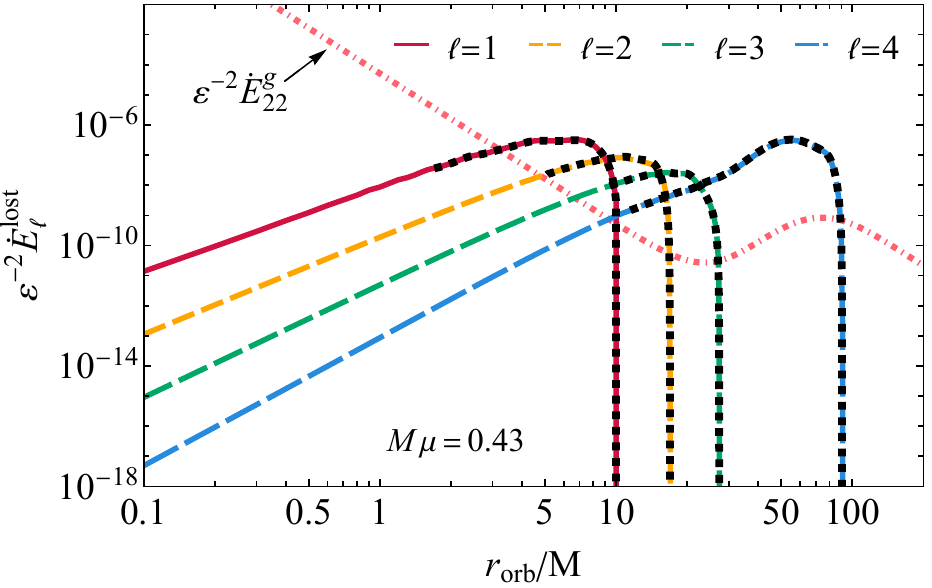}
\caption{Same as Fig.~\ref{Fig:gs-Ls-mu43-plot}, but for the hedgehog NPS configuration.}
\label{Fig:hedge-Ls-mu43-plot}
\end{figure}

The contribution of the four lowest $\ell$ multipoles\footnote{We recall the condition $\ell>0$ for nonvanishing fluxes.} to the total Proca energy loss is shown in Figs.~\ref{Fig:gs-Ls-mu43-plot} and~\ref{Fig:hedge-Ls-mu43-plot}, for the ground-state and hedgehog NPS configurations, respectively. In all cases we show the energy loss summed over $m$-modes, i.e., $\dot{E}^\text{lost}_{\ell} \equiv \sum_{m=1}^\ell \dot{E}^\text{lost}_{\ell m}$ and compare it with the GW flux computed using the leading-order quadrupole formula in the small mass-ratio limit~\cite{Maggiore:2007}:
\begin{equation}
\dot{E}^\text{g}_{22} = \varepsilon^2 \frac{32}{5} \, r_{\text{orb}} \left( \frac{\dd U_n}{\dd r} \right)^3 \bigg|_{r = r_{\text{orb}}} \,.
\end{equation}

For both NPS configurations the GW flux dominates at small orbital radii, namely when $r_{\text{orb}}/M\lesssim 1$ (i.e., $r_{\text{orb}}/M_{\rm BH}\lesssim 50$ for $M_{\text{\tiny BH}}=0.02M$). For $r_{\text{orb}}/M \gtrsim 1$, the total orbital energy loss budget is instead dominated by the Proca energy loss rate up to some $r_{\rm cutoff}$. For $r_{\text{orb}}\gtrsim r_{\rm cutoff}$, the Proca contribution is extremely suppressed and GW fluxes also dominate in this region. We verified that for $\ell>4$ the cutoff radii do not shift appreciably and the total Proca energy loss rates remain qualitatively similar. Hence, restricting our analysis to $\ell\leq 4$ provides a good approximation to the total energy loss budget.

As can be seen in Figs.~\ref{Fig:gs-Ls-mu43-plot} and~\ref{Fig:hedge-Ls-mu43-plot}, $r_{\rm cutoff}\sim 50M$ for the ground-state configuration while $r_{\rm cutoff}\sim 100M$ for the hedgehog configuration, which is compatible with the approximate radius of the corresponding NPSs. The suppression of the Proca losses for $r_{\text{orb}}\gtrsim r_{\rm cutoff}$ is then simple to understand: for $r_{\text{orb}}\gtrsim r_{\rm cutoff}$ the NPS energy density is exponentially small, therefore suppressing Proca energy losses.

It is also worth pointing out that, for the hedgehog NPS, only modes with $\ell+m$ even contribute to the Proca energy loss rate. This follows from the source term $p(r)$ of a point particle in circular, equatorial motion [Eq.~\eqref{eq:source-p}]. Since $p(r)$ is proportional to $Y^{\ell m}(\pi/2,0)$, which vanishes for $\ell+m$ odd, only terms with $\ell+m$ even survive. The emission is further dominated by the $\ell=m$ modes, so it is well approximated by restricting to these modes, as also happens for scalar NBSs~\cite{Duque_2024,Annulli:2020l}. This contrasts with the ground-state NPS, where the coupling between different $\ell$ modes in the master perturbation equations causes all $(\ell,m)$ to contribute to the energy losses. The different $m$-mode contributions are apparent for the $\ell=2$ multipole: in the ground-state case, the contributions from the $(2,1)$ and $(2,2)$ modes appear as two distinct peaks, whereas the hedgehog configuration exhibits a single peak associated with the $(2,2)$ mode (cf. $\ell=2$ curve in Figs.~\ref{Fig:gs-Ls-mu43-plot} and~\ref{Fig:hedge-Ls-mu43-plot}).

The distinct energy density profiles also lead to clear additional differences between the two configurations. For the hedgehog NPS, the dipolar $\ell=1$ mode dominates at small orbital radii $r_{\rm orb}$, whereas for the ground-state solution the dipolar $\ell=1$ and quadrupolar $\ell=2$ modes contribute comparably to the loss rates in the same regime. Moreover, across all modes considered here, the ground-state NPS yields systematically larger Proca energy losses than the hedgehog NPS, with maximum values one to two orders of magnitude higher.

\subsection{Total Proca energy loss rate and comparison with scalar Newtonian boson stars}
\label{subsec:compare_NBS}

Using the results discussed above, we now compute the total Proca energy loss rate obtained by summing over the first four allowed multipoles:
\begin{equation}
    \dot{E}^\text{lost}_{\ell \leq 4} := \sum_{\ell=1}^4 \dot{E}^\text{lost}_{\ell} \,.
    \label{eq:total_flux_All-summed-over}
\end{equation}

\begin{figure}[htb]
\centering
\includegraphics[width=0.48\textwidth,clip]{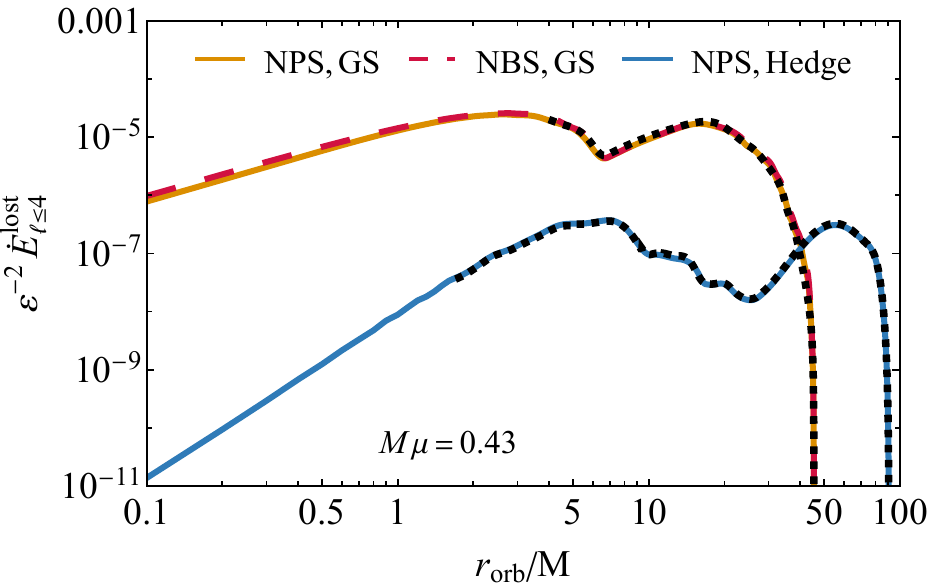}
\caption{Total energy loss rate (considering multipoles up to $\ell=4$) for a perturber in circular motion around three bosonic configurations (see legend; ``GS'' refers to ``ground state'' and ``Hedge'' refers to ``hedgehog''), for $M\mu=0.43$ with a central parasitic BH of mass $M_{\text{\tiny BH}}=0.02M$, obtained using both the source-dominated approximation (colored solid/dashed lines) and the full setup (thick black dashed lines).}
\label{Fig:All-summed-over-plot}     
\end{figure}

Here, we consider the ground-state and hedgehog NPSs, and compare them with the scalar field counterpart, namely the ground-state NBS configuration. In all cases, we again consider a central parasitic BH with mass $M_{\text{\tiny BH}}=0.02M$ and fix $M\mu=0.43$, where $M$ denotes the mass of the NPS or NBS, depending on the case considered.
To compute the scalar energy loss rates in the case of the NBS, we follow Ref.~\cite{Annulli:2020l}. The results are summarized in Fig.~\ref{Fig:All-summed-over-plot}.

As done before, in the presence of a parasitic BH, we only compute the energy loss rates using the full setup in regions where the method provides stable numerical results.
In the region where we are able to compare the total loss rates using the full setup against the ones obtained using the source-dominated approximation, the results are nearly indistinguishable, as expected for dilute bosonic configurations. For the ground-state configurations, we display the loss rates obtained with the full setup only for the NPS, since they essentially overlap with the corresponding NBS results.

Overall, the maximum values for the energy loss rate in the hedgehog NPS background are smaller than those in the ground-state NBS/NPS by roughly one to two orders of magnitude. The energy loss rate in the hedgehog NPS is only larger than in the ground states at orbital radii $r_{\rm orb}\gtrsim 50 M$. This crossover reflects the difference in compactness: the ground state has $R_{98}\sim 50M$ against the hedgehog's $R_{98}\sim 124M$, and since loss rates are suppressed once the orbit lies beyond a configuration's radius, the ground state's rates fall off first. As discussed above, including multipoles beyond $\ell=4$ would not alter these qualitative conclusions.

Interestingly, although the ground-state NBS and NPS differ in several respects~--~scalar versus vector field, spherical versus axially symmetric structure at the field level, and distinct master equations governing the linear perturbations of the background~--~the total energy loss rate in these backgrounds is nearly identical. This similarity can be traced to the structure of the background SP equations. For the ground-state configurations, the SP system takes the same form up to the replacement $h(r)\to \Psi(r)$, where $h(r)$ and $\Psi(r)$ denote the NPS and NBS radial profiles, respectively (cf. Eq.~\eqref{eq:SP-bkg-gs} and Eq.~(39) of Ref.~\cite{Annulli:2020l}). The two configurations therefore share equivalent radial profiles\footnote{For a visual comparison, cf. Fig.~\ref{fig:NPS-plot} and Fig.~2 of Ref.~\cite{Annulli:2020l}.} and, despite the nonspherical nature of the Proca field, the same spherically symmetric energy density distribution. The numerical agreement is nonetheless surprising given the very different master equations governing the linear perturbations. Evaluating the ratio of the total energy loss rates in the ground-state NPS and NBS backgrounds, we find that it oscillates between $\sim 0.8$ and $\sim 1$, with the largest deviations at small $r_{\rm orb}$. While this ratio shows some sensitivity to numerical choices~--~such as the truncation order $L$ and the boundary conditions imposed at the origin and at infinity~--~it varies only mildly under reasonable changes of these parameters and remains generically close to unity, with the loss rate in the NPS ground state being consistently $\sim20\%$ smaller than in the NBS ground state at very small $r_{\rm orb}$. We checked that this conclusion seems to be largely independent on the choice of the free parameters $M\mu$ and $\Mbh /M$.

The main conclusion is therefore that, in the Newtonian regime, the inspiral of a perturber on circular orbits around a ground-state NPS and around a ground-state NBS produces a very similar response in both configurations. By contrast, the total energy losses are clearly more significant in both ground-state configurations than in the hedgehog NPS state. The Newtonian limit is expected to provide a reliable guide for $r_{\rm orb}/M\gg 1$, but a fully relativistic description is ultimately required to assess whether the two ground states can be observationally distinguished. Indeed, at the relativistic level the equivalence between the energy density profiles of the ground-state PS and BS is broken: the PS develops a prolate morphology, unlike the corresponding BS, which remains spherically symmetric~\cite{Herdeiro2023:groundstate,Diez-Tejedor:2026fnc}. We expect these differing morphologies to produce larger discrepancies in the strong-field regime. 

Relativistic studies for scalar BSs clearly indicate that bosonic configurations around supermassive BHs are potentially observable in EMRI waveforms with future detectors such as LISA~\cite{Duque_2024,Khalvati:2024tzz}. This motivates future relativistic studies of EMRIs inside Proca stars, which may be essential to determine whether the properties of the underlying field~--~in particular its intrinsic spin~--~can be measured with such observations.

\section{Conclusions and Outlook}
\label{sec:conclusions}

In this work, we developed a nonrelativistic perturbative framework to study EMRIs into NPSs, with and without a BH at their center.
Previous studies of EMRIs into bosonic compact objects have focused mainly on BSs, both in the Newtonian~\cite{Annulli:2020l,Duque_2024} and relativistic regimes~\cite{Macedo_2013,Guo_2019,Destounis:2023khj,Duque_2024,Macedo:2026}. By contrast, EMRIs into PSs remain largely unexplored. The limited literature connecting EMRIs and Proca fields has instead focused on different physical setups, including EMRIs into BHs with Proca-haired secondaries~\cite{Zi:2024,Zi:2025kerr,Zi:2025qos}, EMRIs into BHs surrounded by superradiant (non self-gravitating) Proca clouds~\cite{Cao:2023fyv,Cao:2024wby}, and EMRIs into static BHs embedded in a cold vector DM environment in which the scattering of the Proca field by the secondary was not modeled~\cite{Karmakar:2025drp}.
The goal of this work was therefore to initiate the study of EMRIs into spherically and axially symmetric PSs in the Newtonian regime, providing a first step toward a fully relativistic description. Our results thus extend the NBS analysis of Ref.~\cite{Annulli:2020l} to its vector counterpart.

We began by constructing the spherical and axisymmetric NPS configurations, reproducing the results in Ref.~\cite{Adshead:2021l}. We confirmed that these correspond to an excited-state and a ground-state solution, respectively, thereby reproducing the nonrelativistic limit of Ref.~\cite{Herdeiro2023:groundstate} (see also~\cite{Diez-Tejedor:2026fnc} for recent comparisons between the relativistic and Newtonian PS solutions).
We then studied the response of these configurations to linear perturbations induced by a point particle in circular, equatorial Keplerian motion around the NPS. For spherically symmetric NPSs, the perturbations decouple into axial and polar parity sectors (Sec.~\ref{subsubsec:hedge-small}). By contrast, for axisymmetric NPSs this separation is lost, leading to couplings between the axial and polar sectors as well as between different multipoles $\ell$ (Sec.~\ref{subsec:gs-NPS-small}). This behavior is expected from analogous studies in general relativity: perturbations around generic spherically symmetric metric backgrounds allow a clean separation into axial and polar sectors (see, e.g., Ref.~\cite{chandrasekhar1998mathematical}), but once spherical symmetry is lost, perturbations can no longer, in general, be separated by parity alone, and different sectors and multipoles may couple. This is analogous to the behavior found when perturbing slowly rotating stars~\cite{kojimaStar,PhysRevD.Kojima} or BHs~\cite{Pani:2012,Pani:2013wsa}.
Finally, in Sec.~\ref{sec:numerical_results} we showed that the rate of energy lost by the point particle through Proca radiation can overcome the energy lost through GW emission in the early inspiral, in agreement with previous work on EMRIs in BSs~\cite{Duque_2024,Macedo:2026}. In the Newtonian regime, we found that a particle on circular orbits around a ground-state NPS and around a ground-state NBS loses a similar amount of energy, different by at most $\sim 20\%$ at small orbital radii, whereas the energy lost in a hedgehog (spherically symmetric) NPS is significantly smaller, for configurations with the same parameters (see Fig.~\ref{Fig:All-summed-over-plot}).

The present work constitutes a first step toward fully understanding the dynamics of EMRIs in the presence of generic forms of ultralight DM. Our results should be a good approximation for dilute bosonic stars, with secondaries moving at sufficiently large orbital radii~\cite{Duque_2024}, and they suggest that, if DM is composed of light boson fields that form bosonic star configurations (in their ground state) at the centers of galaxies, the energy lost by a small body moving inside this environment through scattering of bosons is nearly independent of the intrinsic boson spin. Although this might have been anticipated, given that at the Newtonian level the ground-state NBS and NPS share the same spherically symmetric energy density profile, it is a highly nontrivial result in view of the very different equations that the linear perturbations of these configurations satisfy. It also suggests that the same conclusions should hold for generic (nonequatorial, eccentric) orbits, although this should be confirmed by an explicit computation.

Given these results, the next natural step is to study EMRIs into PSs in a fully relativistic description, a necessary but nontrivial problem owing to the nonspherical configuration of the PS ground state. Such an extension is essential for capturing the strong-field regime and for understanding whether EMRI observations with future detectors such as LISA could actually be used to distinguish between different types of bosonic environments. Since the relativistic PS ground state acquires a prolate morphology, unlike the ground-state BS~\cite{Herdeiro2023:groundstate}, we expect that, once relativistic effects are included, the differences in the energy lost by a point particle moving in a ground-state BS and in a ground-state PS will be larger than what we found here.

Further extensions include adding self-interaction terms, which are known to modify the structure and properties of both BSs and PSs~\cite{Herdeiro:2016gxs,Kling:2017,Guerra:2019,colpi1986boson,Aoki:2022mdn,Minamitsuji:2018kof}, or couplings to additional fields or matter sectors, such as axion-photon or Proca-Higgs-like couplings~\cite{Sanchis_Gual_2022,Brito:2024biy}. Lastly, evidence for chaotic motion has also been found around self-interacting rotating boson stars~\cite{Destounis:2023khj}, and similar results are expected to hold for rotating PSs. We hope to return to some of these issues in future work.

\begin{acknowledgments}
We acknowledge financial support provided by FCT – Fundação para a Ciência e a Tecnologia, I.P., through the ERC-Portugal program Project ``GravNewFields''. We also thank the Fundação para a Ciência e Tecnologia (FCT), Portugal, for the financial support to the Center for Astrophysics and Gravitation (CENTRA/IST/ULisboa) through grant No.~\href{https://doi.org/10.54499/UID/PRR/00099/2025}{UID/PRR/00099/2025} and grant No.~\href{https://doi.org/10.54499/UID/00099/2025}{UID/00099/2025}.
\end{acknowledgments}

\appendix

\section{Schr\"{o}dinger-Poisson system}
\label{app:SP}

In this Appendix, we derive the ``vectorial'' Schr\"odinger-Poisson system [Eqs.~\eqref{eq:SP-system-ps1} and \eqref{eq:SP-system-ps2}] as the Newtonian limit of the Einstein-Proca field equations. Similar calculations can be found in Refs.~\cite{Adshead:2021l,Nambo:2024hao,Diez-Tejedor:2026fnc}, although following slightly different approaches. For clarity and completeness, we restate some of the definitions introduced in the main text and reproduce the relevant equations.

We start from the Einstein-Proca action for a massive complex vector field $\mathcal{A}_\mu$ minimally coupled to gravity,
\begin{equation}
    \mathcal{S}=\int \dd^4 x \sqrt{-g} \left( \frac{R}{16 \pi} -\frac{1}{4} \mathcal{F}_{\mu \nu} \bar{\mathcal{F}}^{\mu \nu} - \frac{1}{2} \mu^2 \mathcal{A}_\mu \bar{\mathcal{A}}^\mu  \right)  \,.
\label{eq:action-EP-app}
\end{equation}
Varying the action with respect to the metric and the Proca field yields the Einstein-Proca system
\begin{equation}
\begin{aligned}
G_{\mu\nu} &= 8 \pi T_{\mu \nu}^{\rm Proca}\,, \\
 \nabla_\mu \mathcal{F}^{\mu \nu} &= \mu^2 \mathcal{A}^\nu \,,
\end{aligned}
\label{eq:EP-system2-app}
\end{equation}
where $T_{\mu \nu}^{\rm Proca}$ is the Proca stress-energy tensor
given in Eq.~\eqref{eq:stress-energy-tensor}. Before taking the Newtonian limit, it is useful to rewrite the Proca field equations in curved spacetime as\footnote{This
form follows from $\nabla_\mu \mathcal{F}^{\mu\nu} = \Box \mathcal{A}^\nu - \nabla^\nu(\nabla_\mu \mathcal{A}^\mu) - R^\nu{}_{\mu} \mathcal{A}^\mu$ together with the Lorenz condition $\nabla_\mu\mathcal A^\mu=0$.}
\begin{equation}
    (\Box - \mu^2) \mathcal{A}^\nu - R^\nu_{\,\,\mu} \mathcal{A}^\mu = 0 \,,
\label{eq:proca-eq-curved}
\end{equation}
where $\Box\equiv\nabla_\mu\nabla^\mu$ is the d'Alembertian operator.

Following Refs.~\cite{Annulli:2020l,Poisson_Will_2014}, we consider the ``Newtonian'' spacetime metric ansatz in harmonic coordinates
\begin{equation}
\begin{aligned}
g_{00} &= -1 - 2U + \mathcal{O}(\varepsilon_{\rm N}^{4})\,,\\
g_{0j} &= \mathcal{O}(\varepsilon_{\rm N}^{3})\,, \quad g_{ij} = (1-2U)\delta_{ij} + \mathcal{O}(\varepsilon_{\rm N}^{4})\,,
\end{aligned}
\label{eq:newtonian-metric-app}
\end{equation}
with $i,j=\{x,y,z\}$ and $\varepsilon_{\rm N}\ll1$ a small scaling parameter. The Newtonian potential
satisfies $U(t,\mathbf{x}) \sim \mathcal{O}(\varepsilon_{\rm N}^2)$, while the terms of order
$\mathcal{O}(\varepsilon_{\rm N}^4)$ in $g_{00}$ and $g_{ij}$, and $\mathcal{O}(\varepsilon_{\rm N}^3)$ in $g_{0j}$ only contribute to post-Newtonian corrections to the field equations, so we neglect them. From Eq.~\eqref{eq:newtonian-metric-app}, the Ricci tensor components are given by
\begin{equation}
\begin{aligned}
R_{00} &= \nabla^{2} U + \mathcal{O}(\varepsilon_{\rm N}^{4})\,,\\
R_{0j} &= \mathcal{O}(\varepsilon_{\rm N}^{3})\,, \quad R_{ij} = \delta_{ij} \nabla^{2} U + \mathcal{O}(\varepsilon_{\rm N}^{4})\,,
\end{aligned}
\label{eq:ricci-app}
\end{equation}
where we consider the nonrelativistic slow motion, $\partial_t \sim \varepsilon_{\rm N} \partial_j$, such that 
\begin{equation}
\partial_j U \sim \mathcal{O}(\varepsilon_{\rm N}^{2})\,, \qquad \partial_t U \sim \mathcal{O}(\varepsilon_{\rm N}^{3})\,.
\end{equation}

We also consider a weak Proca field $|\mathcal{A}_\mu|\ll1$, whose spatial components scale as $\mathcal{A}_j \sim \mathcal{O}(\varepsilon_{\rm N})$. In the nonrelativistic limit, we separate the fast rest-mass oscillation from the slow residual dynamics by writing
\begin{equation}
\mathcal{A}_\mu(t, \mathbf{x}) = \frac{1}{\sqrt{\mu}} \tilde{\mathcal{A}}_\mu(t, \mathbf{x}) e^{-\mathrm{i}\mu t} \,,
\label{eq:ansatz2-appendix}
\end{equation}
where $\tilde{\mathcal{A}}_\mu$ represents an auxiliary vector field, slowly varying on the timescale $\mu^{-1}$. Its spatial components then satisfy\footnote{See Ref.~\cite{Annulli:2020l} for a justification for this scaling.}
\begin{equation}
\tilde{\mathcal{A}}_j \sim \mathcal{O}(\varepsilon_{\rm N})\,, \quad  \partial_i \tilde{\mathcal{A}}_j \sim \mathcal{O}(\varepsilon_{\rm N}^{2})\,, \quad \partial_t \tilde{\mathcal{A}}_j \sim \mathcal{O}(\varepsilon_{\rm N}^{3})\,.
\label{eq:orders-Aj}
\end{equation}
Using the Lorenz condition for the Proca field together with Eq.~\eqref{eq:ansatz2-appendix}, we obtain, at leading order, the constraint
\begin{equation}
    \tilde{\mathcal{A}}_0 \simeq  \frac{\mathrm{i}}{\mu} \partial_j \tilde{\mathcal{A}}^j\,.
\end{equation}
Accordingly, $\tilde{\mathcal{A}}_0\sim\mathcal{O}(\varepsilon_{\rm N}^{2})$. The temporal component is therefore subleading,
$|\tilde{\mathcal{A}}_0| \ll |\tilde{\mathcal{A}}_j|$, and we retain only the spatial components in what follows.

We now consider the spatial components of Eq.~\eqref{eq:proca-eq-curved}
\begin{equation}
   \left( \Box - \mu^2 \right) \mathcal{A}_j-R_{j\mu}\mathcal{A}^\mu=  0 \,.
\label{eq:proca-no-coupling}
\end{equation}
In the leading-order Newtonian approximation adopted here, the gravitational interaction in the term $\Box \mathcal{A}_j$ only enters through the Newtonian potential in the metric component $g_{00}$, with the other metric components only contributing at higher post-Newtonian order. Substituting the ansatz~\eqref{eq:ansatz2-appendix} into
Eq.~\eqref{eq:proca-no-coupling}, and retaining the leading order terms $\mathcal{O}(\varepsilon_{\rm N}^3)$, we obtain
\begin{equation}
    \mathrm{i} \partial_t \tilde{\mathcal{A}}_j = -\frac{1}{2\mu} \nabla^2 \tilde{\mathcal{A}}_j + \mu U \tilde{\mathcal{A}}_j + \frac{1}{2\mu}R_{ji}\tilde{\mathcal{A}}^i\,.
\end{equation}

In order to put this equation in the form of the Schr\"odinger equation, some extra considerations are needed to justify neglecting the curvature-coupling term $R_{ji}\tilde{\mathcal{A}}^i$\footnote{This is unlike what happens in the scalar field case, where no such term is present in the Klein-Gordon equation (see, e.g., Ref.~\cite{Annulli:2020l}).}. This term can be neglected as long as $R_{ji}\tilde{\mathcal{A}}^i/(2\mu^2U\tilde{\mathcal{A}}_{j})\ll 1$. From Eq.~\eqref{eq:ricci-app} we have 
\begin{equation}
\frac{R_{ji}\tilde{\mathcal{A}}^i}{2\mu^2U\tilde{\mathcal{A}_{j}}}\simeq \frac{\nabla^2U}{2\mu^2U} \sim \frac{k^2 U}{2\mu^2 U}\sim \frac{k^2}{2\mu^2}\,,  
\end{equation}
where $k^{-1}\sim R$ is the typical  lengthscale of the solutions of interest. The condition $k^2/(2\mu^2)\ll 1$ has a clear physical interpretation: it is simply stating that the vector bosons' kinetic energy should be much smaller than their rest-mass energy, as is required in the nonrelativistic limit we are considering. For the NPS solutions we can take $R$ to be their typical size,\footnote{The exact choice for how $R$ is defined is not important for our argument, but one can for example define $R$ as the radius that encloses $98\%$ of the NPS mass, as done in the main text.} and so the condition becomes $2R\mu\gg 1$, i.e., the size of the objects should be much larger than the Compton wavelength of the vector boson. For all cases presented in the main text, this condition is always satisfied. Therefore, for our purposes, we can neglect the curvature-coupling terms, and the spin-1 Schr\"odinger equation becomes
\begin{equation}
    \mathrm{i} \partial_t \tilde{\mathcal{A}}_j = -\frac{1}{2\mu} \nabla^2 \tilde{\mathcal{A}}_j + \mu U \tilde{\mathcal{A}}_j\,.
\end{equation}

Since the Newtonian limit of the Einstein field equations is well established in the literature (see, e.g., Refs.~\cite{Misner:1973,Carroll:2004,Schutz:1985}), we do not rederive it here. For the metric ansatz~\eqref{eq:newtonian-metric-app}, the $00$-component of the Einstein tensor is, at leading order,
$G_{00}  = 2 \nabla^2U$. Consequently, the $00$-component of the Einstein field equations in Eq.~\eqref{eq:EP-system2-app} results in
\begin{equation}
    \nabla^2 U = 4\pi T_{00}^{\rm Proca}\,,
\label{eq:poisson-appendix}
\end{equation}
where the $00$-component of the
Proca stress-energy tensor is given by
\begin{align}
\begin{split}
    T^\text{Proca}_{00} =& -\mathcal{F}_{\sigma0}\bar{\mathcal{F}}_{0}{}^{\sigma}-\frac{1}{4} g_{00} \mathcal{F}_{\sigma \rho} \bar{\mathcal{F}}^{\sigma \rho} \\
    &+ \mu^2 \left[ \mathcal{A}_{0} \bar{\mathcal{A}}_{0}-\frac{1}{2} g_{00} \mathcal{A}_\sigma \bar{\mathcal{A}}^\sigma \right] \,.
\label{eq:stress-energy-tensor-00}
\end{split}
\end{align}

Introducing the electric- and magnetic-type components $\mathcal{E}_i \equiv \mathcal{F}_{0i}$ and $\mathcal{B}_{ij} \equiv \mathcal{F}_{ij}$, Eq.~\eqref{eq:stress-energy-tensor-00} can be written, at leading order, as
\begin{equation}
\begin{aligned}
T_{00}^{\rm Proca}
&\simeq
\frac{1}{2}\mathcal{E}_i\bar{\mathcal{E}}^{i}
+\frac14(1+2U)\mathcal{B}_{ij}\bar{\mathcal{B}}^{ij} \\
&+\frac{\mu^2}{2}\left(\mathcal{A}_0\bar{\mathcal{A}}_0+\mathcal{A}_j\bar{\mathcal{A}}^{j}\right) 
+\mu^2 U \mathcal{A}_{j}\bar{\mathcal{A}}^{j}\,.
\end{aligned}
\label{eq:T00_31_split}
\end{equation}

In the weak-field, nonrelativistic limit, we have $|\mathcal{A}_0|\ll|\mathcal{A}_j|$ and $|\partial_i \mathcal{A}_j| \ll |\mathcal{A}_j|$. Hence, the temporal component
$\mathcal{A}_0$ and the magnetic-type components $\mathcal{F}_{ij}$ are
subleading. Substituting the ansatz~\eqref{eq:ansatz2-appendix} into
Eq.~\eqref{eq:T00_31_split} and retaining only the leading order terms, we obtain
\begin{equation}
    T^{\rm Proca}_{00} = \mu \tilde{\mathcal{A}}_{j} (\tilde{\mathcal{A}}^{j})^* + \mathcal{O}(\varepsilon_{\rm N}^4)\,.
\end{equation}
Using Eq.~\eqref{eq:poisson-appendix}, we obtain the Poisson equation, at leading order $\mathcal{O}(\varepsilon_{\rm N}^2)$,
\begin{equation}
    \nabla^2 U = 4\pi\mu \tilde{\mathcal{A}}_{j}(\tilde{\mathcal{A}}^{j})^*\,.
\end{equation}
Combining this result with the spin-1 Schr\"odinger equation derived previously, the Einstein-Proca system reduces, in the Newtonian and weak-field limit, to the Schr\"odinger-Poisson system for $\tilde{\mathcal{A}}_j$ and $U$.

\section{Proca energy and Noether charge fluxes radiated to infinity}
\label{app:Erad-Qrad-flux}

In this Appendix, we derive the expressions used in the main text for the Proca energy and Noether charge fluxes radiated to infinity. We work directly in the asymptotic region, where the spacetime metric in spherical coordinates reduces to
$\eta_{\sigma\rho}=\diag(-1,1,r^2,r^2\sin^2\theta)$, and retain only the leading order contributions $\mathcal{O}(\varepsilon^2)$.

The energy radiated per unit time at infinity, $\dot E^{\rm rad}$, is given by Eq.~\eqref{eq:E-rad-Proca}. From the perturbed Proca stress-energy tensor, Eq.~\eqref{eq:stress-energy-tensor-pert}, the relevant $rt$-component at leading order takes the form
\begin{align}
\begin{split}
    \delta T^\Proca_{r t} (r \to \infty) &\sim - \eta^{\sigma \rho}\delta\mathcal{F}_{\sigma(r}\delta\bar{\mathcal{F}}_{t)\rho} + \mu^2  \delta\mathcal{A}_{(r} \delta\bar{\mathcal{A}}_{t)} \,,
\end{split}\label{eq:stress-energy-tensor-pert-rt}
\end{align}
where $\delta \mathcal{F}_{\mu\nu}=\partial_\mu \delta \mathcal{A}_\nu - \partial_\nu \delta \mathcal{A}_\mu$. The spatial components of the perturbed Proca field, $\delta\mathcal{A}_j$, are those defined in Eq.~\eqref{eq:fourier-pert-VSH}, while the temporal component $\delta\mathcal{A}_t$ is obtained from the asymptotic Lorenz condition $\partial_\mu  \,\delta\mathcal{A}^\mu \approx 0$.

Using the symmetry relation $\mathcal{Z}_1(\omega,\ell,m;r) = (-1)^m \mathcal{Z}_2(-\omega,\ell,-m;r)^*$, where $\mathcal{Z}_k\equiv \{Z_k,Z_{k\mathcal{P}},Z_{k\mathcal{A}} \}$ for $k=1,2$, together with the spherical harmonic identity $\left( Y^{\ell, m} \right)^* = (-1)^m Y^{\ell,-m}$, the spatial components can be written entirely in terms of $\mathcal{Z}_1$, 
\begin{equation}
    \delta \mathcal{A}_j = \frac{2}{\sqrt{2\pi}}\sum_{\ell,m} \int \dd\omega\, \delta \mathbf{A}_{(1) j}^{\omega \ell m} e^{-\mathrm{i} \omega t}  e^{-\mathrm{i}\Omega t} \,,
    \label{eq:proca-pert-spatial-app}
\end{equation}
where $\Omega=\mu-\gamma$ is defined in Sec.~\ref{sec:NPS-bkg}.

At spatial infinity, the polar and axial radial functions satisfy the outgoing Sommerfeld radiation condition,
\begin{equation}
        \mathcal{Z}_{1}^{\omega \ell m}({r \to \infty}) \sim \mathcal{Z}_{1}^{\infty, \omega\ell m} \,e^{\mathrm{i} k_1 r} r^{\mathrm{i} \frac{  M \mu^2}{k_1} }\,,
        \label{eq:boundary-Zs-rinf-app}
\end{equation}
where $\mathcal{Z}_{1}^{\infty, \omega\ell m}$ are the asymptotic amplitudes and $k_1\equiv\sqrt{(\omega+\Omega)^2-\mu^2}$ is the corresponding wave number. For a point particle on a circular, equatorial orbit, the Dirac delta $\delta(\omega-m\omega_{\rm orb})$ in Eq.~\eqref{eq:source-p} restricts the source contribution to the discrete frequencies $\omega=m\omega_{\rm orb}$.

Accordingly, we factor out the Dirac delta from the asymptotic amplitudes as $\mathcal{Z}_{1}^{\infty , \omega\ell m} \equiv \tilde{\mathcal{Z}}_{1}^{\infty, \ell m} \delta(\omega - m\omega_{\rm orb})$ and define $\tilde{\mathcal{Z}}_{1}^{\infty, \ell m} \equiv \tilde{\mathcal{Z}}_{1}^{\infty,\ell m} (m\omega_{\rm orb};r_{\rm orb})$. The frequency integral in Eq.~\eqref{eq:proca-pert-spatial-app} then collapses to
$\omega=m\omega_{\rm orb}$, and the four components of the perturbed Proca field take the form
\begin{equation}
    \delta \mathcal{A}_\mu = \frac{2}{\sqrt{2\pi}}\sum_{\ell,m}  \delta \tilde{\mathcal{A}}_{\mu} e^{-\mathrm{i} m\omega_{\rm orb} t}  e^{-\mathrm{i}\Omega t}\,,
    \label{eq:ansatz-pert-Proca-hat-1}
\end{equation}
where
\begin{equation}
\begin{split}
    \delta \tilde{\mathcal{A}}_t &= - \frac{\tilde{k}_1}{m\omega_{\rm orb}+\Omega} \frac{\tilde{Z}_{1}^{\infty, \ell m}}{r}\,Y^{\ell m} e^{\mathrm{i} \tilde{k}_1 r} r^{\mathrm{i} \frac{ M \mu^2}{\tilde{k}_1} } \,,\\
    \delta \tilde{\mathcal{A}}_r &=  \frac{\tilde{Z}_{1}^{\infty, \ell m}}{r}\,Y^{\ell m} e^{\mathrm{i} \tilde{k}_1 r} r^{\mathrm{i} \frac{ M \mu^2}{\tilde{k}_1} }   \,,\\
    \delta \tilde{\mathcal{A}}_\theta &=   \left( \tilde{Z}^{\infty, \ell m}_{1 \mathcal{P}} \,Y_{,\theta}^{\ell m} + \tilde{Z}^{\infty, \ell m}_{1 \mathcal{A}} \,\frac{ Y_{,\varphi}^{\ell m}}{\sin\theta} \right) e^{\mathrm{i} \tilde{k}_1 r} r^{\mathrm{i} \frac{ M \mu^2}{\tilde{k}_1} }\,, \\
    \delta \tilde{\mathcal{A}}_\varphi &=   \Bigl( \tilde{Z}^{\infty, \ell m}_{1\mathcal{P}} \,Y_{,\varphi}^{\ell m} - \tilde{Z}^{\infty, \ell m}_{1\mathcal{A}} \,Y_{,\theta}^{\ell m} \sin\theta \Bigr) e^{\mathrm{i} \tilde{k}_1 r} r^{\mathrm{i} \frac{ M \mu^2}{\tilde{k}_1} } \,.
\end{split}\label{eq:ansatz-pert-Proca-hat}
\end{equation}
Here, we used the shorthand notation $Y_{,\theta}\equiv \partial_{\theta} Y$ and  $Y_{,\varphi}\equiv \partial_{\varphi}Y$, and defined $\tilde{k}_1\equiv\sqrt{(m\omega_{\rm orb}+\Omega)^2-\mu^2}$.

From Eqs.~\eqref{eq:E-rad-Proca}, \eqref{eq:stress-energy-tensor-pert-rt}, and \eqref{eq:ansatz-pert-Proca-hat-1}, together with the relevant scalar spherical harmonic identities, we obtain the Proca energy flux radiated to infinity:
\begin{equation}
\begin{aligned}
\dot E^{\rm rad}
=&\frac{2}{\pi}\sum_{ \ell, \ell',m}
\Re\!\left[\sqrt{\left(m\omega_{\rm orb}+\Omega\right)^2-\mu^2}\right]
\left(m\omega_{\rm orb}+\Omega\right)
\\
\times&\Bigg\{\,
\tilde Z_{1\mathcal P}^{\infty,\ell' m}\left(\tilde Z_{1\mathcal P}^{\infty,\ell m}\right)^{*}\, I^{\mathcal{P}}_{\bar\Omega} 
+
\tilde Z_{1\mathcal A}^{\infty,\ell' m}\left(\tilde Z_{1\mathcal A}^{\infty,\ell m}\right)^{*}
\, I^{\mathcal{A}}_{\rm \bar\Omega} 
\\
&+\left[
\tilde Z_{1\mathcal P}^{\infty,\ell' m}\left(\tilde Z_{1\mathcal 
A}^{\infty,\ell m}\right)^{*}
-
\tilde Z_{1\mathcal A}^{\infty,\ell' m}\left(\tilde Z_{1\mathcal P}^{\infty,\ell m}\right)^{*}
\right]\, I^{\mathcal{\times}}_{\bar\Omega}
\\
&+\frac{\mu^{2}\,\tilde Z_{1}^{\infty,\ell' m}\left(\tilde Z_{1}^{\infty,\ell m}\right)^{*}}
{\left(m\omega_{\rm orb}+\Omega\right)^{2}}\,I_{\bar\Omega} \Bigg\} \,.
\end{aligned}
\label{eq:Eloss-expression2-hat}
\end{equation}
In Eq.~\eqref{eq:Eloss-expression2-hat}, the selection rule $m=m'$, which follows from the orthogonality relations below, has already been imposed for compactness. The terms $I_{\bar\Omega}$, $I^{\mathcal{P}}_{\bar\Omega}$, $I^{\mathcal{A}}_{\bar\Omega}$, $I^{\times}_{\bar{\Omega}}$ are integrals over the sphere involving spherical harmonics which, following the conventions of Sec.~15.2 in Ref.~\cite{Ferrari:2020}, satisfy the orthogonality relations,
\begin{align}
I_{\bar\Omega}&\equiv\bigl\langle Y^{\ell m},Y^{\ell' m'}\bigr\rangle
=\delta^{\ell\ell'}\delta^{mm'}\,, \label{eq:Y-ortho}\\
I^{\mathcal{P}}_{\bar\Omega}&\equiv\bigl\langle \gamma^{ab}\mathbf{Y}^{\ell m}_{a},\mathbf{Y}^{\ell' m'}_{b}\bigr\rangle
=\ell(\ell+1)\delta^{\ell\ell'}\delta^{mm'}\,, \label{eq:VSHY-ortho}\\
I^{\mathcal{A}}_{\bar\Omega}&\equiv\bigl\langle \gamma^{ab}\mathbf{S}^{\ell m}_{a},\mathbf{S}^{\ell' m'}_{b}\bigr\rangle
=\ell(\ell+1)\delta^{\ell\ell'}\delta^{mm'}\,, \label{eq:VSHS-ortho}\\
I^{\times}_{\bar\Omega}&\equiv\bigl\langle \gamma^{ab}\mathbf{Y}^{\ell m}_{a},\mathbf{S}^{\ell' m'}_{b}\bigr\rangle=0
\,, \label{eq:VSH-cross}
\end{align}
where $\gamma_{ab}=\diag(1,\sin^2\theta)$ is the metric on the two-sphere $\mathbf{S}^2$, and
$\langle f,g\rangle\equiv\int \dd\theta \dd\varphi \sin\theta\,f^*g$ denotes the corresponding scalar product.

In the nonrelativistic regime, $\gamma\ll\mu$ and $\omega_{\rm orb}\ll\mu$, the wave number reduces to $\tilde{k}_1=\sqrt{(m\omega_{\rm orb}+\Omega)^2-\mu^2} \simeq \sqrt{2\mu(m\omega_{\rm orb}-\gamma)}$. Using the orthogonality relations in Eqs.~\eqref{eq:Y-ortho}--\eqref{eq:VSH-cross} and the nonrelativistic limit of the wave number, Eq.~\eqref{eq:Eloss-expression2-hat} reduces to
\begin{equation}
    \begin{aligned}
        &\dot{E}^\text{rad} = \frac{2}{\pi} \sum_{\ell,m} \Re \left[ \sqrt{2\mu(m \omega_{\rm orb} - \gamma)} \right]   \left(\mu-\gamma+m\omega_{\rm orb}\right)  \times\\
     &   \!\Bigg\{\!\ell (\ell+1) \left(|\tilde Z_{1\mathcal{P}}^{\infty,\ell m}|^2+|\tilde Z_{1\mathcal{A}}^{\infty,\ell m}|^2\right)  + \frac{|\tilde Z_1^{\infty,\ell m}|^2 \mu^2}{(\mu-\gamma+m\omega_{\rm orb})^2} \!\Bigg\}
     \label{eq:Erad-expression2-hat}\,,
    \end{aligned}
\end{equation}
which corresponds to Eq.~\eqref{eq:Erad-expression2} in the main text. Notice that in the main text, we suppressed the mode labels $(\ell,m)$ and the tildes on the asymptotic amplitudes $\tilde{\mathcal Z}_1^{\infty,\ell m}$ for simplicity. Up to normalization conventions, this Eq.~\eqref{eq:Erad-expression2-hat} agrees with the corresponding result derived in Appendix~A of Ref.~\cite{Zi:2024}, when taking the Newtonian limit.

The corresponding Noether charge flux is obtained by applying the same procedure to Eq.~\eqref{eq:Q-flux-Proca}. At leading order $\mathcal{O}(\varepsilon^2)$, the relevant $r$-component of the perturbed Noether current, Eq.~\eqref{eq:noether-charge-pert}, reads
\begin{equation}
    \delta j_r (r \to \infty) \sim - \frac{\mathrm{i}}{2} \eta^{\sigma \rho} \left[ \delta\bar{\mathcal{F}}_{r \rho} \,\delta \mathcal{A}_\sigma - \delta \mathcal{F}_{r \rho} \,\delta \bar{\mathcal{A}}_\sigma \right]\,. 
\label{eq:noether-charge-pert-r}
\end{equation}
Inserting the ansatz~\eqref{eq:ansatz-pert-Proca-hat-1} in Eq.~\eqref{eq:Q-flux-Proca}, and using again the orthogonality relations~\eqref{eq:Y-ortho}--\eqref{eq:VSH-cross} and the nonrelativistic wave number approximation, we obtain the Noether charge flux radiated to infinity:
\begin{equation}
    \begin{aligned}
        &\dot{Q}^\text{rad} = -\frac{2}{\pi} \sum_{\ell,m} \Re \left[ \sqrt{2\mu(m \omega_{\rm orb} - \gamma)} \right]   \times\\
     &   \Bigg\{ \ell (\ell+1) \left(|\tilde Z_{1\mathcal{P}}^{\infty,\ell m}|^2+|\tilde Z_{1\mathcal{A}}^{\infty,\ell m}|^2\right)  + \frac{|\tilde Z_1^{\infty,\ell m}|^2 \mu^2}{(\mu-\gamma+m\omega_{\rm orb})^2} \Bigg\}\,.
    \end{aligned}
    \label{eq:Qrad-expression2-hat}
\end{equation}
This expression corresponds to Eq.~\eqref{eq:Qrad-expression2} in the main text.

\section{Convergence tests}
\label{sec:convergence}

In this Appendix we study the numerical convergence of the energy loss rates in the NPS ground-state background presented in Sec.~\ref{sec:numerical_results}, when increasing the truncation order $L$ discussed in Sec.~\ref{subsec:NPS-gs}.

\subsubsection{Full-setup method}

\begin{figure}[htb]
\includegraphics[width=0.45\textwidth]{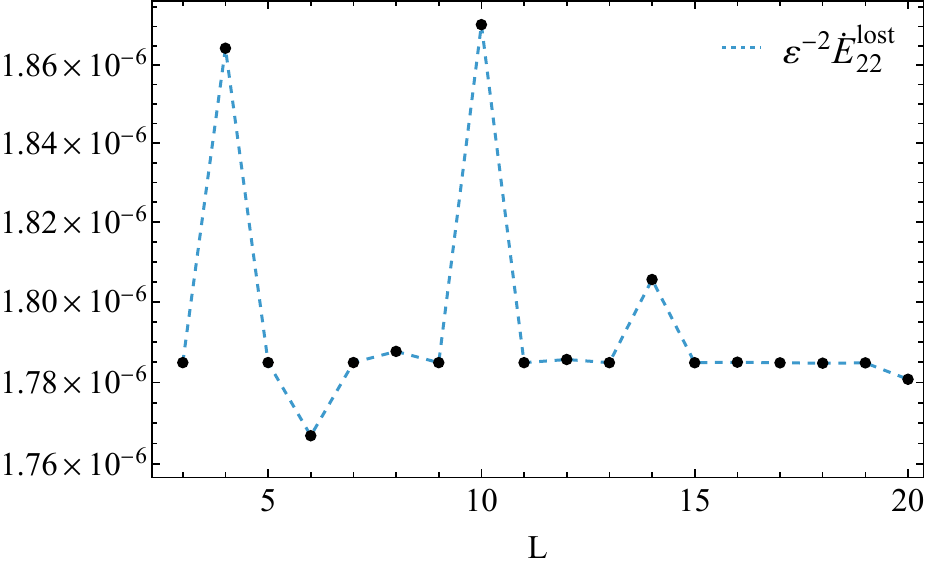}
\caption{Energy loss rate $\varepsilon^{-2} \dot{E}^{\rm lost}_{22}$ as a function of the multipolar truncation order $L$, for a point particle at fixed orbital radius $r_{\rm orb}/M=5$ around the ground-state NPS with a central parasitic BH of mass $M_\text{\tiny BH}=0.02M$ and coupling $M\mu=0.43$. The results are obtained using the full-setup method.}
\label{Fig:convergence_rorbs_FunMatrix}
\end{figure}

Let us first look at the results obtained with the full setup. We start by fixing an orbital radius at which the inverse of the fundamental matrix is well behaved for all cases considered. We choose $r_{\text{orb}}/M=5$ and, for a given azimuthal number $m$ and multipole $\ell$, we increase the multipolar truncation order $L$. Following Sec.~\ref{subsec:with_a_parasitic_BH}, we take a ground-state NPS with a central parasitic BH of mass $M_{\text{\tiny BH}}=0.02M$ and coupling $M\mu=0.43$. Although the convergence behavior with increasing $L$ is qualitatively similar across all modes considered, in Fig.~\ref{Fig:convergence_rorbs_FunMatrix} we show the energy loss rate for the $(\ell,m)=(2,2)$ mode, considering a truncation order up to $L=20$. Due to the structure of the master equations, at the first possible truncation, $L=m$, the fluxes vanish if $m$ even and are nonzero for $m$ odd. As such, to obtain nonvanishing fluxes for $m=2$, the lowest truncation we can use is $L=3$.

As the truncation $L$ is increased, the loss rate displays an oscillatory convergence pattern, with a few pronounced peaks at low truncation orders, before approaching an effective plateau for $L\gtrsim 16$. Taking the plateau value as a reference, the largest relative deviation is $\Delta\dot E^{\rm lost}_{22}/\dot E^{\rm lost}_{22}\sim 0.05$. A similar pattern is found for other values of $(\ell,m)$ and $r_{\rm orb}/M$. Extending the truncation further could reveal additional oscillations, but we expect them to remain within a comparably small relative deviation. The computational cost increases rapidly with $L$, through the fundamental matrix dimensionality ${\rm dim}=14(L-m)+10$ (e.g., ${\rm dim}\ge206$ for $L\ge16$), making it impractical to consider very large truncation orders. However, Fig.~\ref{Fig:convergence_rorbs_FunMatrix} shows that even at the lowest possible truncation order ($L=3$) we already obtain very good estimates for the loss rates. Therefore, we chose the following truncation orders:
\begin{enumerate}[label=(\roman*)]
    \item For the results shown in Sec.~\ref{subsec:w_wo_BH_numerical-results}, namely the computation of the mode $\ell=m=2$, we use $L=4$.
    \item For the results shown in Sec.~\ref{subsec:with_a_parasitic_BH}, we use $L=m+4$ for a given $m$, keeping the fundamental matrix dimension fixed at ${\rm dim}=66$ for that particular $m$ mode.
\end{enumerate}

These choices were based on finding a reasonable compromise between convergence and computational cost. Larger truncations leave the loss rates nearly unchanged while substantially increasing the runtime.

\subsubsection{Source-dominated approximation}
\label{app:conv-source-dominated}

\begin{figure}[htb]
\includegraphics[width=0.45\textwidth]{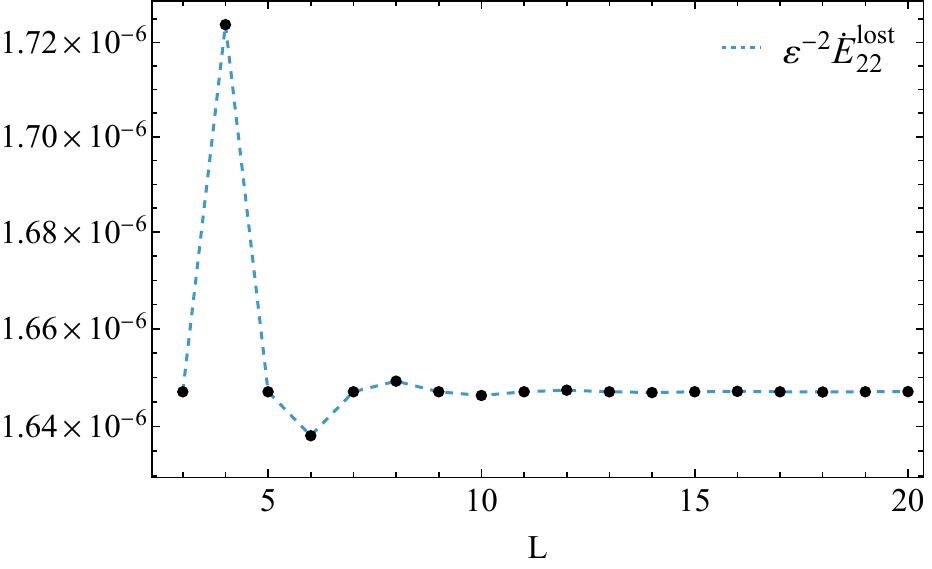}
\caption{Same as Fig.~\ref{Fig:convergence_rorbs_FunMatrix}, but using the source-dominated approximation.}
\label{Fig:convergence_rorbs_source_m2l2}
\end{figure}

For the convergence analysis in the source-dominated approximation we follow the same procedure as above. The results in this case are shown in Fig.~\ref{Fig:convergence_rorbs_source_m2l2}. The trend is similar to what we found when using the full setup, although for this case there is a clear convergence for $L\geq 10$. Therefore, for the results shown in Sec.~\ref{sec:numerical_results}, when considering the source-dominated approximation, we used as truncation $L=10$.

\section{Comparison between methods}
\label{subsec:3_methods}

In this Appendix, we compare the three approaches discussed in Sec.~\ref{sec:methods}: the full-setup method, without further approximations; the source-dominated approximation, where we compute the perturbations to the gravitational potential $\delta U$ assuming $\nabla^2 \delta U \simeq 4\pi P$; and the high-frequency approximation, which builds upon the source-dominated approximation and further assumes $\omega_{\text{orb}} \gg \gamma, \,\mu U(r)$.

The high-frequency approximation is only strictly valid when we include a parasitic BH (see discussion in Sec.~\ref{sec:numerical_results}). We follow the setup of Sec.~\ref{subsec:w_wo_BH_numerical-results}, including a central parasitic BH for both the ground-state and hedgehog NPSs, so that the high-frequency approximation can be used, and focus on the mode $\ell=m=2$.

\begin{figure}[htb]
\includegraphics[width=0.45\textwidth]{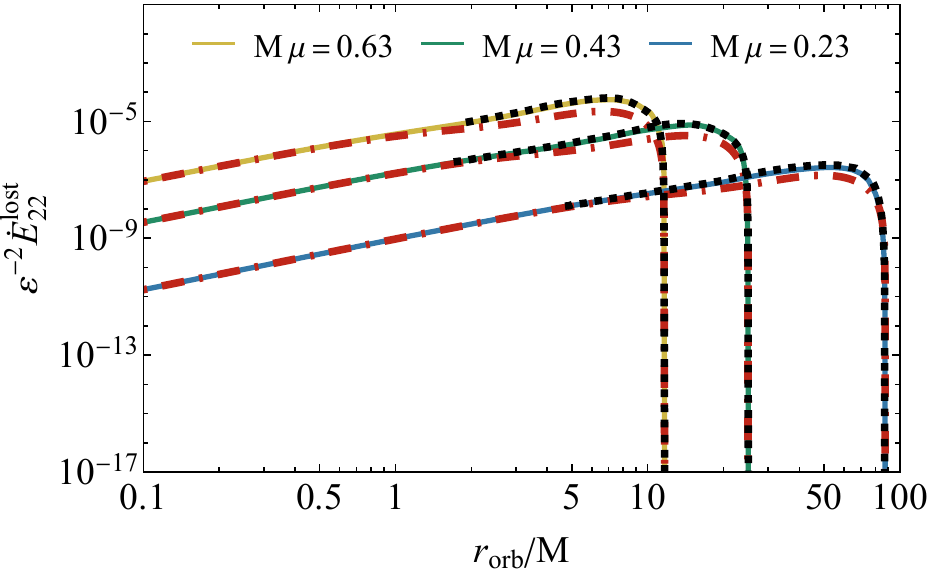}
\caption{Ground-state NPS. Energy loss fluxes for the $\ell=m=2$ mode, computed with the three methods: full setup (black dashed lines), source-dominated approximation (colored solid lines), and high-frequency approximation (red dot-dashed lines). We consider a parasitic BH of mass $M_\text{\tiny BH}=0.02M$, adopting truncation orders $L=4$ for the full setup and $L=10$ for the source-dominated and high-frequency approximations.}
\label{Fig:Fun_An_ND_GS}
\end{figure}

\begin{figure}[htb]
\includegraphics[width=0.45\textwidth]{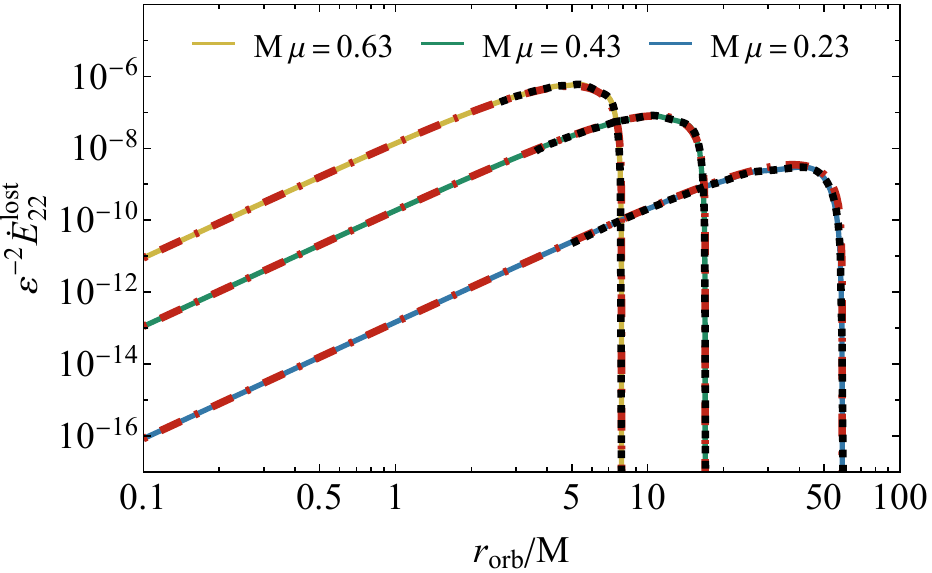}
\caption{Hedgehog NPS. Energy loss fluxes for the $\ell=m=2$ mode, computed with the three methods: full setup (black dashed lines), source-dominated approximation (colored solid lines), and high-frequency approximation (red dot-dashed lines); with a parasitic BH of mass $M_\text{\tiny BH}=0.02M$.}
\label{Fig:Fun_An_ND_hedge}
\end{figure}

Figures~\ref{Fig:Fun_An_ND_GS} and \ref{Fig:Fun_An_ND_hedge} show the energy loss rates computed with the three methods. For both the ground-state and hedgehog configurations, the source-dominated and high-frequency approximations produce nearly identical results at small orbital radii. Since the matrix inversion in the full-setup method becomes numerically ill defined for $r_\text{orb}/M\ll1$, no comparisons can be made for small radii with that method. However, for the ground state without a parasitic BH, for which we can use the full setup at smaller $r_\text{orb}$, we verify that the full-setup method overlaps with the source-dominated approximation (see Fig.~\ref{Fig:gs-Mus-l2m2-plot}). 

Moreover, as stated in Sec.~\ref{sec:numerical_results}, the high-frequency approximation is in principle only valid at sufficiently small orbital radii and becomes increasingly inaccurate at large distances. In particular, for the ground-state NPS, both conditions $\omega_{\text{orb}} \gg \gamma$ and  $\omega_{\text{orb}}\gg \mu U(r)$ are only strictly valid when $r_{\rm orb}/M\ll 1,3,10$ for $M\mu=0.63,0.43,0.23$, respectively, explaining the differences between the three methods seen at large orbital radii in Fig.~\ref{Fig:Fun_An_ND_GS}. Interestingly, however, for the hedgehog NPS, the three methods show a very good agreement across all the allowed orbital radii, even though the high-frequency conditions $\omega_{\text{orb}} \gg \gamma, \mu U(r)$, are only strictly satisfied when $r_{\rm orb}/M\ll 3,6,20$ for $M\mu=0.63,0.43,0.23$, respectively. 

In contrast, both the full-setup and the source-dominated approximations remain valid throughout all possible orbital radii, with both methods producing similar results in regions where we are able to use the full setup. It should be noted, however, that for the ground-state NPS, the source-dominated approximation consistently gives slightly lower loss rates when compared with the full setup. On the other hand, for the hedgehog configuration, up to numerical accuracy, both methods give the same results. This is consistent with the fact that the ground-state configuration is more compact than the hedgehog one, and therefore the source-dominated approximation is expected to work slightly better for the hedgehog configuration.

\bibliography{ref.bib}

\end{document}